\PassOptionsToPackage{hidelinks}{hyperref}
\documentclass[aps,prd,twocolumn,superscriptaddress, nofootinbib]{revtex4-2}

\usepackage[utf8]{inputenc}
\usepackage{amsmath}
\usepackage{amsfonts}
\usepackage{physics}
\usepackage{comment}
\usepackage{bm}
\usepackage[separate-uncertainty=true]{siunitx}
\usepackage{graphicx}
\usepackage{orcidlink}
\usepackage{natbib}
\usepackage[hidelinks]{hyperref}
\usepackage{mathtools}
\usepackage{xcolor}
\usepackage{ctable}
\usepackage[normalem]{ulem}
\usepackage{multirow}
\usepackage[export]{adjustbox}
\usepackage[bottom]{footmisc}

\makeatletter
  \renewcommand\p@subsection{}
  \renewcommand\p@subsubsection{}
\makeatother

\newcommand{\rmd}{{\rm d}}

\newcommand{\chieff}{{\chi_{\rm eff}}}
\newcommand{\cumchidiff}{{C_{\rm diff}}}
\newcommand{\mchirp}{{\mathcal{M}}}

\newcommand{\phiref}{\phi_{\rm ref}}

\newcommand{\los}{\bm{\hat n}}

\newcommand{\phijlhat}{\hat\phi_{JL}}
\newcommand{\thetajn}{\theta_{JN}}

\newcommand{\thetaint}{\bm{\theta}_{\rm int}}
\newcommand{\thetaext}{\bm{\theta}_{\rm ext}}

\begin{document}
\allowdisplaybreaks
\title{Sampler-free gravitational wave inference using matrix multiplication}

\author{Jonathan Mushkin\,\orcidlink{0009-0009-9057-9581}}
\email{jonathan.mushkin@weizmann.ac.il}
\affiliation{\mbox{Department of Particle Physics \& Astrophysics, Weizmann Institute of Science, Rehovot 76100, Israel}}
\author{Javier Roulet\,\orcidlink{0000-0003-3268-4796}}
\affiliation{\mbox{TAPIR, Walter Burke Institute for Theoretical Physics, California Institute of Technology, Pasadena, CA 91125, USA}}
\author{Barak Zackay\,\orcidlink{0000-0001-5162-9501}}
\affiliation{\mbox{Department of Particle Physics \& Astrophysics, Weizmann Institute of Science, Rehovot 76100, Israel}}
\author{Tejaswi Venumadhav\,\orcidlink{0000-0002-1661-2138}}
\affiliation{\mbox{Department of Physics, University of California, Santa Barbara, CA 93106, USA}}
\affiliation{\mbox{International Centre for Theoretical Sciences, Tata Institute of Fundamental Research, Bangalore 560089, India}}
\author{Oryna~Ivashtenko\,\orcidlink{0000-0002-2805-9405}}
\affiliation{\mbox{Department of Particle Physics \& Astrophysics, Weizmann Institute of Science, Rehovot 76100, Israel}}
\author{Digvijay Wadekar\,\orcidlink{0000-0002-2544-7533}}
\affiliation{\mbox{Department of Physics and Astronomy, Johns Hopkins University,
3400 N. Charles Street, Baltimore, Maryland, 21218, USA}}
\affiliation{\mbox{School of Natural Sciences, Institute for Advanced Study, 1 Einstein Drive, Princeton, NJ 08540, USA}}
\author{Ajit Kumar Mehta\,\orcidlink{0000-0002-7351-6724}}
\affiliation{\mbox{Department of Physics, University of California, Santa Barbara, CA 93106, USA}}
\author{Matias Zaldarriaga\,\orcidlink{0009-0007-8315-6703
}}
\affiliation{\mbox{School of Natural Sciences, Institute for Advanced Study, 1 Einstein Drive, Princeton, NJ 08540, USA}}

\date{\today}

\begin{abstract}
    Parameter estimation (PE) for compact binary coalescence (CBC) events observed by gravitational wave (GW) laser interferometers is a core task in GW astrophysics. We present a method to compute the posterior distribution efficiently without relying on stochastic samplers. First, we show how to select sets of intrinsic and extrinsic parameters that efficiently cover the relevant phase space. We then show how to compute the likelihood for all combinations of these parameters using dot products. We describe how to assess and tune the integration accuracy, making the outcome predictable and adaptable to different applications. The low computational cost allows full PE in minutes on a single CPU, with the potential for further acceleration using multiple CPUs or GPUs. We implement this method in the \texttt{dot-PE}\footnote{\url{https://github.com/JonathanMushkin/dot-PE}} package, enabling sensitive searches using the full evidence integral for precessing CBCs and supporting large waveform banks ($\sim10^5$--$10^6$ waveforms), regardless of waveform generation cost.
\end{abstract}

\maketitle

\section{Introduction}
\label{sec:intro}
Compact objects (neutron stars and black holes) in binaries undergoing mergers emit gravitational waves (GW), and are being frequently detected by ground-based laser interferometers \cite{2015CQGra..32g4001L, Aasi2012}. The intricate information about the masses of the compact objects and their spin vectors is encoded in the fine details of the received gravitational waves. As of the third observing run (O3) of the LIGO-Virgo collaboration, $\sim100$ compact binary coalescence (CBC) events have been confirmed by search pipelines maintained by various groups \cite{GWOSC2023, Wadekar2023, Abbott2023_GWTC3, Nitz4OGC2023}, while $\sim200$ more are expected from the fourth observing run \cite{CapoteO4performance2025, LIGO2025O4Milestone}. As the detectors' sensitivity increases and more detectors are commissioned, the detection rate is expected to grow rapidly \cite{Abac2025ET,Evans2021, LIGO-T2200287,Saleem2022LIGOindia}. 

Within the Bayesian parameter estimation framework, the process of inferring the physical properties of the emitting systems from the GW events is computationally expensive. The main challenge is to efficiently explore the 15-dimensional parameter space of the waveform. Numerous sampling techniques and codes have been developed to address it (including, but not limited to, \cite{Farr2014, Farr2014PhRvD, Veitch2015, Pankow2015, Zackay2018, Biwer2019, Dax2021,Gabbard2021, Islam2022, Rose2022, Fairhurst2023, Wong2023, roulet2024extrinsic, Nitz2024arXiv}). For a recent review, see \cite{Roulet2024}. 

In this work, we propose a novel technique to evaluate the posterior distribution, which provides both samples from the posterior and integrates it for the Bayesian evidence, requiring only a few minutes on a single CPU-core of a commercial computer available at the time of writing (2025). We implement the method in \texttt{dot-PE}, a publicly available pure \texttt{Python} package. The summary of the method, along with a breakdown of this paper, is provided in the following section.

\section{Summary of this work}
\begin{figure*}
    \centering
    \includegraphics[width=\linewidth]{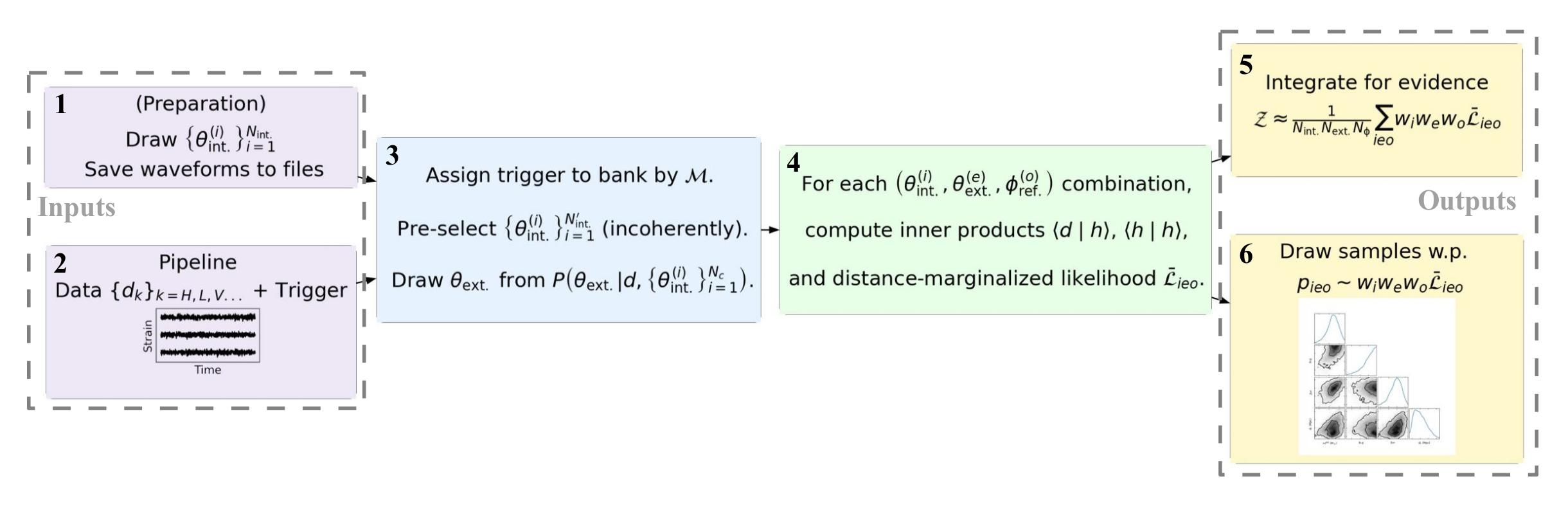}
    \caption[]{Flowchart of \texttt{dot-PE}. Intrinsic waveform bank, (block 1, Section~\ref{subsec:intrinsic_samples}) is performed once per mass range and can be re-used for different events. Block 2 is the output of a conventional search pipeline: detector strains and trigger information. In block 3, the phase space is reduced. The trigger is assigned to a bank by its $\mchirp$. Intrinsic samples are accepted based on their incoherent maximum-likelihood (Section~\ref{subsec:preselection}), reducing the samples from $N_{\rm int.}$ to $N_{\rm int.}'$ (see Section~\ref{subsec:integration} and Fig.~\ref{fig:selection_and_ESS} for the relation between $N_{\rm int.}'$ and the expected performance of the integration). Extrinsic samples are drawn using adaptive-importance sampling  conditioned on the data and a subset of $N_c$ intrinsic samples (Section~\ref{subsec:extrinsic_samples}). In block 4, the intrinsic and extrinsic components are combined to evaluate the inner products and likelihoods, per intrinsic-extrinsic-phase combination (Section~\ref{subsec:lnlike_as_matmul}). The outputs of the method are the evidence integral (block 5) and samples drawn from the posterior distribution (block 6).}
    \label{fig:block_diagram}
\end{figure*}
In this work, we propose a new scheme for the Bayesian inference problem of CBC using GW interferometer data, including the effects of higher modes and precession. A block diagram of our method is presented in Fig.~\ref{fig:block_diagram}. The waveform, and therefore also the likelihood, can be written as a sum-of-products of components, that depend on separate sets of parameters. Schematically, it is written as 
\begin{equation}
    \bm \theta = (\thetaint, \thetaext, \phi_{\rm ref.}, d_L)
\end{equation}
and
\begin{equation}
    \ln\mathcal{L}({\bm \theta}) = \sum_{f} A_f(\thetaint) B_{f}(\thetaext)C_{f}(\phiref) E_{f}(d_L)
\end{equation}
where $\thetaint$ are the intrinsic parameters (which set the waveforms themselves), $\thetaext$ are the extrinsic parameters (which set the projection of the waveforms only to detector data), and $\phiref$ is the reference orbital phase, $d_L$ is the (luminosity) distance. The details of the parameters and the components will explained in the body of the text.\footnote{The summation over azimuthal modes, polarizations and detectors are omitted here. They are included in the body of the text.}

Traditional matched-filtering pipelines (e.g. cite \cite{Allen2012, Usman2016, Messick2017, Venumadhav2019})
search for candidate signals by performing matched filtering (cross-correlation) of the detector strain against a bank of template waveforms. Times at which a particular template yields a large signal-to-noise (SNR) 
value are reported as triggers. Given such a trigger, the goal of this work is to evaluate the likelihood with sufficient 
density across the entire parameter space. Naively, the prior ranges are too 
wide to be covered with a reasonable number of samples. We therefore narrow 
the parameter space using the following steps: 
\begin{enumerate}
\item The pipeline trigger provides a rough estimate of the chirp mass, which lets us restrict the intrinsic samples to the corresponding pre-computed set 
of waveforms (see Sec.~\ref{subsec:intrinsic_samples}). For example, while the 
full parameter space may cover chirp masses from $3$ to $300\,{\rm M}_\odot$, a trigger of a template 
with chirp mass near $25\,{\rm M}_\odot$ allows us to focus on the much narrower range 
around $20$–$30\,{\rm M}_\odot$.
\item
Given the data, we include or exclude each of the $N_{\rm int.}$ intrinsic
waveforms according to a likelihood maximized \emph{incoherently} across detectors
(i.e., the maximization over $\theta_{\rm ext}$, $\phi_{\rm ref}$, and $d_L$ is done
per detector, without cross-detector consistency). This pre-selection retains
$N_{\rm int.}'$ intrinsic samples (Sec.~\ref{subsec:preselection}).
\item Given a random selection of intrinsic samples, we employ an adaptive importance sampling method \cite{roulet2024extrinsic} to draw extrinsic parameters that are consistent with the arrival times at the different detectors.
\end{enumerate}
After the phase space narrowing stage, the parameter space is represented by the Cartesian product of the three selected sets: 
\begin{equation*}
    \left\{\thetaint^{(i)}\right\}_{i=1}^{N_{\rm int.}'}, \left\{\thetaext^{(e)}\right\}_{e=1}^{N_{\rm ext.}}, \left\{\phiref^{(o)}\right\}_{o=1}^{N_{\phi}}.
\end{equation*}
The luminosity-distance $d_L$ will be marginalized over.
Once the components for each point are calculated ($A_{if} = A_f\big(\thetaint^{(i)}\big)$, $B_{ef} = B_f\big(\thetaext^{(e)}\big)$,and $C_{of} = C_f\big(\phiref^{(o)}\big)$), the distance-marginalized likelihood $\overline{\mathcal{L}}$ \textit{at all possible combinations} can be evaluated through a series of matrix multiplications, and a distance marginalization function~$F$:
\begin{align*}
    \ln\overline{\mathcal{L}}_{ieo} &= \ln\overline{\mathcal{L}}\left(\thetaint^{(i)},\thetaext^{(e)},\phiref^{(o)}\right) \\
    &= F\left(\sum_{f} A_{if} B_{ef} C_{of}\right)
\end{align*}
Matrix operations are far more efficient than serialized computations both due to algorithmic and implementation considerations. The distance-marginalized likelihoods can then be averaged to evaluate the Bayesian evidence (Bayes factor):
\begin{equation}
\label{eq:Z_sum}
    \mathcal{Z} \approx\frac{1}{N_{\rm int.}N_{\rm ext.}N_{\rm \phi}} \sum_{ieo} w_{i}w_{e}w_{o}\overline{\mathcal{L}}_{ieo}
\end{equation}
with $w$ being the probabilistic weights of the samples, related to the drawing process.
The applications of this method are numerous: 
\begin{enumerate}
    \item Detection: The main motivation is to compute $\mathcal{Z}$, which is the 
    optimal test statistic for a search \cite{NP_lemma}, quickly and reliably, as 
    a second stage after a matched-filtering pipeline.
    \item Parameter estimation: Detected events are analyzed by drawing samples from the posterior distribution of the source parameters, defined by the probabilistic weights in Eq.~\eqref{eq:Z_sum}. By modifying the weights $w$ in post-processing, we can also resample from different priors.
    \item The individual maximum-likelihood point can be used for fitting and signal consistency checks~\cite{Allen2005PhysRevD}. 
    \item Fast sampling from the posterior can be used for quick sky-localization, which is useful of electromagnetic followup.
\end{enumerate}  
The method presented here carries two important advantages with respect to traditional sampling methods:
\begin{itemize}
    \item There is little-to-no online waveform generation in this method, as discussed in Section~\ref{subsec:intrinsic_samples}. This allows the use of more accurate waveform models, which would substantially slow down a traditional stochastic sampler (see Section~\ref{subsec:bayesian_inference}).
    \item The runtime and accuracy of the method can be predicted by the number of samples used (demonstrated in Appendix~\ref{app:complexity}), allowing us to tune the input sizes and bank characteristics according to the task in hand.
\end{itemize}

The method stands on two pillars: one is the ability to decompose the likelihood into components and perform matrix evaluations and marginalization efficiently (explained in Sections~\ref{subsec:waveform}--\ref{subsec:lnlike_as_matmul}). The second is the ability to efficiently select points from the smaller parameter spaces (explained in Sections~\ref{subsec:intrinsic_samples}--\ref{subsec:extrinsic_samples}).

\begin{figure*}
\includegraphics[width=\linewidth]{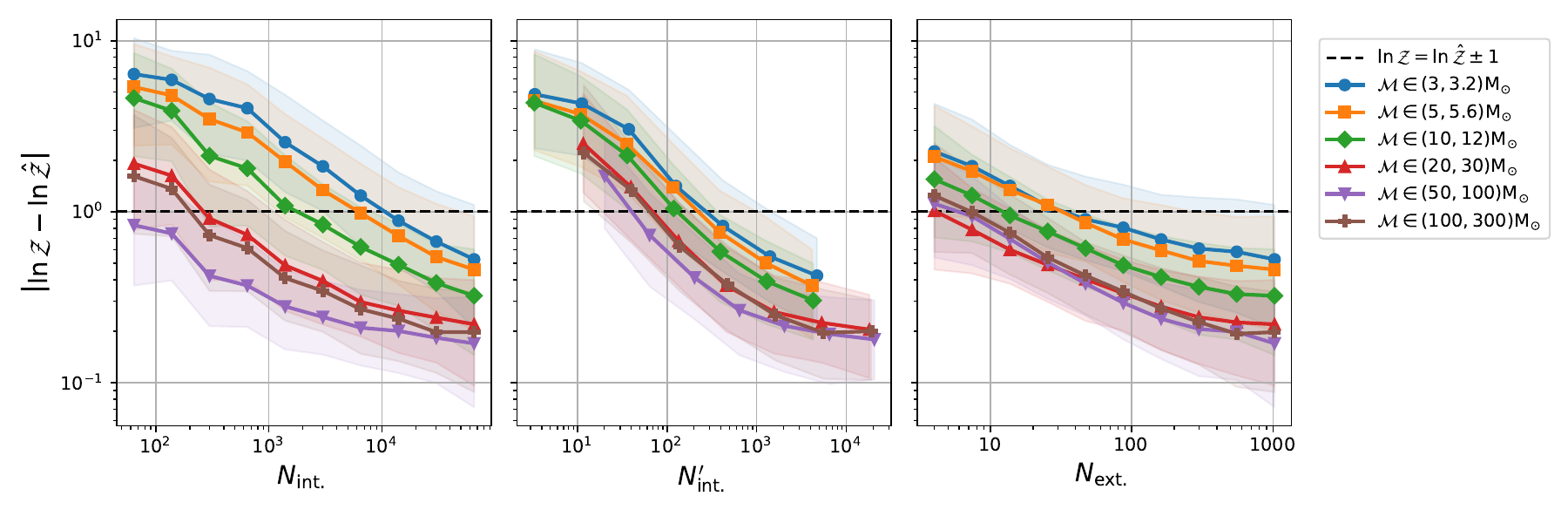}
\caption[]{Convergence of the evidence integral across intrinsic waveform banks with varying mass ranges (Section~\ref{subsec:intrinsic_samples}), shown in different colors and markers as indicated in the top right legend.
\textbf{Left:} Absolute value of the relative error $|\ln\mathcal{Z}-\ln\hat{\mathcal{Z}}|$ for $2^{10} = 1024$ injections per mass range, with $N_{\rm int.}$ from $2^6$ to $2^{16} \approx 6.5 \times 10^4$, and fixed $N_{\rm ext.} = 2^{10}$. Reference values $\hat{\mathcal{Z}}$ are computed with much finer banks of $N_{\rm int.} = 2^{18} \approx 2.6 \times 10^5$. Solid lines show medians; shaded bands span 25th–75th percentiles. The error decreases with $N_{\rm int.}$ until saturation due to extrinsic sample count and $\phiref$ resolution.
\textbf{Center:} Same as left, but plotted against $N_{\rm int.}'$, the number of intrinsic samples retained after pre-selection (Section~\ref{subsec:preselection}). The improved alignment between mass ranges—relative to the left panel—illustrates that convergence is largely governed by the number of relevant samples near the posterior support.
\textbf{Right:} Absolute value of the relative error versus $N_{\rm ext.}$, ranging from $2$ to $2^{10}$, for fixed $N_{\rm int.} = 2^{16}$.}
\label{fig:convergence}
\end{figure*}

\section{Preliminaries}
\label{sec:preliminaries}
In this section we will review the basics for Bayesian inference and integration of the evidence. The review is not exhaustive, and limits the discussion to what is relevant in this work, in addition to the current community-standard in the field of parameter estimation. A recent more detailed review of the topic, in the context of GW astronomy, can be found in \cite{Roulet2024}.
\subsection{Bayesian Inference}
\label{subsec:bayesian_inference}
In the Bayesian inference framework, the likelihood $p(d\vert H,\boldsymbol{\theta})$ is the probability of observing the data $d$ under the assumption of hypothesis $H$ with specific parameters $\boldsymbol{\theta}$. The prior probability distribution $\pi(\boldsymbol{\theta})$, the posterior distribution $p(\boldsymbol{\theta}|d)$ and the likelihood are related through Bayes' theorem. 
\begin{equation}
    p(\boldsymbol{\theta}|H,d) = \frac{p(d|H,\boldsymbol{\theta})\pi(H,\boldsymbol{\theta})}{\int p(d|H,\boldsymbol{\theta}')\pi(H,\boldsymbol{\theta}') {\rm d}\boldsymbol{\theta}'},
\end{equation}
where the denominator is the evidence, or probability to observe the data given the hypothesis $H$. Our goal in this work is to evaluate the evidence ratio, or Bayes factor $\mathcal{Z}$:
\begin{equation}
\mathcal{Z} = \frac{\int p(d|H_1,\boldsymbol{\theta})\pi(\boldsymbol{\theta} | H_1){\rm d}\boldsymbol{\theta}}{\int p(d|H_0,\boldsymbol{\theta}')\pi(\boldsymbol{\theta}'| H_0){\rm d}\boldsymbol{\theta}'}
\end{equation}
In the context of this work, $H_0$ is the noise hypothesis, which states that the data at detector $k$ is a stationary Gaussian noise $n_k$ with auto-correlations that are expressed through the one-sided frequency-domain power-spectral density $S_k(f)$ per detector $k$:
\begin{equation}
    H_0:\quad d_k(f) = n_k(f).
\end{equation} 
The competing hypothesis $H_1$ is the signal hypothesis, which states that there is a gravitational wave signal from a compact binary coalescence embedded in the noise, with specific parameters $\boldsymbol{\theta}$:
\begin{equation}
    H_1:\quad d_k(f) = n_k(f) + h_k(f;\boldsymbol{\theta})
\end{equation}
where the exact dependence of the signal at each detector on the parameters is discussed in Section~\ref{subsec:waveform}. Comparing the two hypotheses, the normalization of the noise probability density function cancels out, as the terms are independent of the parameters. It is therefore useful to define the likelihood-ratio $\mathcal{L}(\boldsymbol{\theta})$ 
\begin{equation}
    \mathcal{L}(\boldsymbol{\theta}) = \frac{p(d|H_1,\boldsymbol{\theta})}{p(d|H_0)},
\end{equation}
the log of which is given as a subtraction between two inner products:
\begin{equation}
\label{eq:lnlike}
    \ln\mathcal{L}({\bm \theta}) = \Re\left\langle d \mid h({\bm \theta}) \right\rangle - \frac{1}{2}\langle h\left({\bm \theta}\right) \mid h\left({\bm \theta}\right) \rangle.
\end{equation}
The inner product $\langle \cdot \mid \cdot \rangle$ is the sum of detector-wise frequency integral, approximated by a sum:
\begin{align}
\label{eq:inner_product}
    \langle a \mid b\rangle &= 4 \sum_k \int_{0}^{\infty}\frac{a_k(f)b_k^\ast(f)}{S_k(f)}\rmd f \\
    &\approx 4\Delta f\sum_k\sum_{f_{\rm min}}^{f_{\rm max}} \frac{a_k(f)b_k^\ast(f)}{S_k(f)}
\end{align}
where $f_{\rm min}$, $f_{\rm max}$ and $\Delta f$ are set by the characteristics of the detector and the duration of the data segments. In practice we use relative-binning (heterodyning) to compute the inner products; see Appendix~\ref{app:relative_binning}. The evidence ratio $\mathcal{Z}$ is the prior-weighted integral of the likelihood ratio:
\begin{equation}
\label{eq:Z}
    \mathcal{Z} =\int \pi({\bm \theta})\mathcal{L}({\bm \theta}) {\rm d}{\bm \theta}.
\end{equation}

\subsection{Integration methods}
\label{subsec:integration_methods}
The fastest and most accurate way to integrate $\mathcal{Z}$ is analytical. Unfortunately, analytical integration is rarely possible in Bayesian inference problems. If the dimensionality of $\boldsymbol{\theta}$ is low (say unidimensional), it may be numerically evaluated on a grid, as the Riemann sum:
\begin{equation}
    \mathcal{Z} \approx \sum_{i=1}^{N} \pi(\theta_i)\mathcal{L}(\theta_i)(\theta_i-\theta_{i-1}). 
\end{equation}
or higher-order methods. This approach scales badly with the dimension of the parameter space, as the error $\Delta\mathcal{Z}$ scales like the spacing $\Delta \theta$. Assuming equal spacing in all dimensions, $\Delta\mathcal{Z}\sim N^{-1/{\rm dim.}}$. At higher dimension, an ensemble average, or Monte-Carlo (MC) integration, is a more efficient way to evaluate $\mathcal{Z}$, with independent identically-distributed (i.i.d) samples drawn from $\pi(\boldsymbol{\theta})$:
\begin{equation}
\label{eq:mc_integral}
    \mathcal{Z}\approx \frac{1}{N}\sum_{i=1}^{N}\mathcal{L}(\boldsymbol{\theta}_i),\quad \boldsymbol{\theta}_i\sim \pi(\boldsymbol{\theta})\, \text{ i.i.d}.
\end{equation}
Since the samples are independent, the estimation variance scales like $N^{-1}$, or $\Delta\mathcal{Z}\sim N^{-1/2}$. If it is possible to estimate where are the more contributing regions to the integral in the parameter space, expressed as a fiducial prior $\pi'(\boldsymbol{\theta})$, the samples can be drawn from it and re-weighted to correct for the over- and under-densities. This is known as importance sampling (IS) \cite{Owen2013}:
\begin{equation}
\label{eq:IS_integral}
    \mathcal{Z}\approx \frac{1}{N}\sum_{i=1}^{N}\frac{\pi(\boldsymbol{\theta}_i)}{\pi'(\boldsymbol{\theta}_i)}\mathcal{L}(\boldsymbol{\theta}_i),\quad \boldsymbol{\theta}_i\sim \pi'(\boldsymbol{\theta})\, \text{ i.i.d}.
\end{equation}
While it retains the same scaling $\Delta\mathcal{Z}\sim N^{-1/2}$, the pre-factor can be made smaller by proper selection of $\pi'(\boldsymbol{\theta})$. A drawback of MC integration is that i.i.d.\ samples create random fluctuations in density, which in turn translate to fluctuations in the estimate. This can be mitigated using quasi-random samples (or quasi-Monte Carlo, QMC, sequence) \cite{Owen2013}. Said sequences attempt to minimize the discrepancy, which is the difference between the fractional number of samples within phase-space volume element and the one expected from the prior. The error scaling of QMC sequences (namely the Halton sequence, used here) depends on the dimensionality of the problem \cite{Morokoff1995}, but it typically outperforms MC samples and can reach $\Delta \mathcal{Z}\sim N^{-1}$.

The samples drawn in either MC or QMC using Eq.~\eqref{eq:mc_integral} or \eqref{eq:IS_integral} carry their probabilistic weights $p_n$, in the form of the summands, 
\begin{equation}
    p_n = \mathcal{L}(\theta_n)
\end{equation}
or 
\begin{equation}
\label{eq:IS_prob_weights}
    p_n = \frac{\pi(\theta_n)\mathcal{L}(\theta_n)}{\pi'(\theta_n)},
\end{equation}
respectively. The probabilistic weights enable us to resample, that is, to draw from the $N$ samples according to the probabilistic weights $\left\{ p_n \right\}_{n=1}^{N}$. The samples can be used to describe the posterior, visualize it using one- and two-dimensional histograms, and compute integrals under the posterior distribution, such as mean values, variances, and the probabilities of lying above or below particular values. The distribution of the weights quantifies the confidence in the evidence integral estimate, through the \textit{effective sample size} \cite{kong1992note, liu1995blind}:
\begin{equation}
\label{eq:n_effective}
    N_{\rm eff.} = \frac{\left(\sum_{n=1}^{N} p_n\right)^2}{\sum_{n=1}^N p_n^2}
\end{equation}
which satisfies $1\leq N_{\rm eff.}\leq N$. After computing $\{p_n\}_{n=1}^N$, the effective sample size can indicate whether the evidence estimate is dominated by a single or a few samples, making it prone to large variation, or comprises many similar contributions, making it more reliable \footnote{While the effective sample size is justified for i.i.d samples, we expect the same qualitative behavior in QMC samples.}.

If the volume of the parameter space is too large, in the sense of number of dimensions and/or extent of the dimensions (relative to the relevant resolution required), (Q)MC with or without IS could fail to produce many samples in the relevant regions. More adaptive methods are then employed, such as Markov chain Monte Carlo (MCMC) \cite{1953JChPh..21.1087M,Hastings1970} or nested sampling (NS) \cite{skilling2006nested, Buchner2023Nested}, which iteratively suggest and accept random points in the phase space based on the likelihood of the previous point(s). The random sampling algorithms provide samples drawn from the posterior and the evidence integral. Though MCMC does not directly evaluate $\mathcal{Z}$, some adaptations do allow it, see \cite{Gelman1998}.

A dominant failure mode of both MCMC and NS is multimodality, i.e.\ the presence
of disjoint posterior peaks in the parameter space. Even though modern
implementations attempt to mitigate this issue, it remains a limiting factor
for both methods \cite{Speagle2020, Buchner2023Nested, Ashton2022}.

\section{Method}
\subsection{Waveform Decomposition}
\label{subsec:waveform}
The imprint of the gravitational waves from CBC sources into the strain at the detector depends on 15 parameters. We limit to discussion to quasi-circular (zero eccentricity) binaries without tidal deformability. We separate the parameters into two kinds: the \textit{intrinsic} parameters, which are the two component masses ($m_1,m_2$), two component dimensionless spins (${\bm \chi}_{1,2} = {\bm S}_{1,2}/m_{1,2}^2$), and the inclination angle, or the angle between the line-of-sight and the orbital angular momentum ($\cos\iota = \hat{\bm n} \cdot \hat{\bm{L}}$)
\begin{equation}
    \label{eq:thetaint}  \thetaint= \left(m_1,\, m_2,\, \bm \chi_1,\, \bm \chi_2,\, \iota\right)
\end{equation}
These parameters set the observed waveform. The masses and spins sets the evolution of the binary according to the equation of motion. We also group the inclination with the intrinsic parameters, as its evolution under spin-induced orbital precession and its imprint on the waveform are non-trivial to track analytically \cite{Apostolatos1994, Schmidt2012}. These effects, governed by the time-dependent Euler angles, are incorporated in dedicated waveform generation codes, e.g. \cite{Pratten2021, Yu2023, Ossokine2020PhysRevD}.
The values of the time-evolving parameters $\iota$, $\bm{\chi}_1$ and $\bm \chi_2$ are specified at a reference time or orbital frequency. As a whole, $\thetaint$ sets the intrinsic waveform $h(f;\thetaint)$, for each azimuthal mode $m$ and polarization $p\in\{+,\times\}$, fixing the rest of the parameters to their default values: 
\begin{equation}
    h_{mp}(f;\thetaint) = h_{mp}(f;\thetaint,d_L{=}d_L^0, \phiref{=}0,\psi{=}0)
\end{equation}
The remaining 6 parameters are \textit{extrinsic}, they set the position, orientation and timing of the GW event relative to the detectors. Their effect on the observed signals is easy to account for without the need for waveform generation codes, and their priors are independent of the intrinsic parameters. We will separate them into four categories, according to their functional contribution to the waveform, discussed below. The largest subset will be called the extrinsic parameters:
\begin{equation}
\label{eq:thetaext} \thetaext =\left(\lambda,\, \varphi,\, \psi,\, t_\oplus\right)
\end{equation}
and it includes the sky-position of the source relative to the Earth, expressed as latitude $\varphi$ and longitude $\lambda$ (i.e., adopting a rotating-sky frame with fixed Earth, as opposed to right ascension $\alpha$ and declination $\delta$); the arrival time of the source to the center of the Earth $t_\oplus$; and the polarization angle $\psi$. $\thetaext$ will set the detector $k$'s response to wave polarization $p$, $F_{kp}(\lambda,\varphi,\psi)$:
\begin{align}
    \label{eq:response_plus}
    F_{k,+} &= \mathcal{R}_k(\lambda,\varphi) \cos(2(\psi-\psi_k^0))\\
    \label{eq:response_cross} 
    F_{k,\times} &= \mathcal{R}_k(\lambda,\varphi)\sin(2(\psi-\psi_k^0))
\end{align}
where $\mathcal{R}_k$ is set by the waves' propagation direction relative to the plane containing the detector arms, and $\psi^0_k$ is determined by the arms' orientation within the plane \cite{Whelan2013}. The time of arrival at detector $k$ is set by the sky location of the binary $\hat{\bm n}(\lambda,\varphi)$ and position of the detector ${\bm r}_k$, given that waves propagate at the speed of light $c$: 
\begin{equation}
\label{eq:t_at_detector}
 t_k(\thetaext) = t_{\oplus} + {\bm r}_k \cdot \hat{\bm n}(\lambda, \varphi)/c.
\end{equation}

Time-shifts manifest as multiplication by a frequency-dependent phase:
\begin{equation}
    T(f;\thetaext) = e^{-i2\pi f\, t_k(\thetaext)}
\end{equation}

The two remaining parameters are treated independently. The reference-phase $\phiref$ is the orbital phase at the reference orbital frequency. In the spherical harmonic decomposition, each frequency domain azimuthal mode $m$ is multiplied by a phase factor 
\begin{equation}
    \Phi_m(\phiref) = e^{im\phiref}.
\end{equation}

Finally, the wave amplitude scales inversely with the luminosity distance $d_L$:
\begin{equation}
\label{eq:distance_scaling}
    h(\thetaint,\thetaext,d_L) = h(\thetaint,\thetaext,d_L^0) \frac{d_L^0}{d_L}
\end{equation}

Combining all the effects, the strain due to the gravitational waves, as measured at the detector, is 
\begin{multline}
\label{eq:h_k}
h_k(f; \thetaint, \thetaext, \phiref, d_L) 
\\= \frac{d_L^0}{d_L} \sum_{m,p} h_{mp}(f; \thetaint) \, F_{kp}(\thetaext)
T_k(f; \thetaext) \, \Phi_m(\phiref)
\end{multline}
This decomposition is well known and utilized \cite{Pankow2015, Singer2016, Islam2022, Tiwari2023PhysRevD, roulet2024extrinsic}. We note that the separation assumes the Earth does not rotate and $F_{+,\times}$ remain constant throughout the duration in which the signal resides in the detectors' sensitive band. If both $h(t)$ and $F_{kp}(t)$ are time-series, their time-domain product does not convert to a frequency-domain product. This assumption would break down in next-generation detectors, where the lower frequencies (which correspond to longer durations) will be accessible \cite{Abac2025ET,Chan2018PhRvD}.

\subsection{Likelihood components as matrix products}
\label{subsec:lnlike_as_matmul}
Matrix multiplications are faster than serial evaluations both due to the algorithms (see \cite{strassen1969gaussian} for the renowned algorithm with computational complexity $\mathcal{O}(n^{\log_2 7})$), and due to code implementation (including the memory handling and communication costs, see chapter 1.6 of \cite{bjorck2015numerical}). This holds to the point where it is favorable to perform orders of magnitude more operations (including many low $\overline{\mathcal{L}}$ evaluations) though matrix operations instead of performing a smaller number of serial evaluations. Also, this method is trivially parallelizable and could benefit from speed of graphics processing units (GPUs).

Given sets of points from the three subspaces of $\thetaext$, $\thetaint$ and $\phiref$, we can compute the two inner products in \eqref{eq:lnlike} for some fixed luminosity distance $d_L^0$, we can evaluate the best-fit (over $d_L$) likelihood, and also find summary statistics sufficient for marginalization. Given the points 
\begin{equation*}
    \left\{\thetaint^{(i)}\right\}_{i=1}^{N_{\rm int.}'}, \left\{\thetaext^{(e)}\right\}_{e=1}^{N_{\rm ext.}}, \left\{\phiref^{(o)}\right\}_{o=1}^{N_{\phi}}.
\end{equation*}
we evaluate the components:
\begin{align}
h_{impf} &= h_{mp}\left(f; \thetaint^{(i)}\right), \\
F_{ekp} &= F_{kp}\left(\thetaext^{(e)}\right), \\ 
\Phi_{om} &= \Phi_m\left(\phiref^{(o)}\right)=e^{im\phiref^{(o)}}, \\
\Phi_{omm'} & = \Phi_{om} \Phi_{om'}^{\ast} = e^{i(m-m')\phiref^{(o)}},\\
T_{ekf} &= T_k\left(f; \thetaext^{(e)}\right)=e^{-2i\pi f t_k \left(\thetaext^{(e)}\right)}.
\end{align}
Given said components, the inner products that will comprise $\ln\mathcal{L}$ at some selection of $i,e,o$ at a given reference $d_L^0$, taken to be 1Mpc in this work, are
\begin{equation}
\label{eq:dh_ieo}
    \langle d | h\rangle_{ieo} = \Re \sum_{m=1}^{l_{\rm max}}\sum_{p=+,\times}\sum_{k}\langle d_k| h_{impf} T_{ekf}\rangle F_{ekp} \Phi_{om}^\ast
\end{equation}
and
\begin{equation}
\label{eq:hh_ieo}
\begin{split}
    \langle h | h\rangle_{ieo} & =\\ &\sum_{m,p}\sum_{m',p'}\sum_k
    \langle  h_{impf}| h_{im'p'f} \rangle
    F_{ekp}F_{ekp'}\Phi_{omm'}
\end{split}
\end{equation}
Given the values of Eqs.~\eqref{eq:dh_ieo}--\eqref{eq:hh_ieo} at $d_L^0$, the likelihood at some arbitrary $d_L$ is
\begin{equation}
    \label{eq:lnlike_ieo_arb_dL}
    \ln\mathcal{L}_{ieo}(d_L) = \frac{\langle d \mid h\rangle_{ieo}}{\left(d_L/d_L^0\right)} - \frac{1}{2}\frac{\langle h \mid h \rangle_{ieo}}{\left(d_L/d_L^0\right)^2} 
\end{equation}
One can maximize Eq.~\eqref{eq:lnlike_ieo_arb_dL} over $d_L$ for the best-fit or maximum-likelihood (ML) $\ln\mathcal{L}_{ieo}$:
\begin{equation}
    \label{eq:bestfit_lnl}
    \ln\mathcal{L}_{ieo}^{\rm ML} = 
    \begin{cases}
    \dfrac{\left|\langle d \mid h \rangle_{ieo}\right|^2}{2\langle h \mid h \rangle_{ieo}} & \langle d \mid h \rangle_{ieo} > 0 \\
    0 & \langle d \mid h \rangle_{ieo} \leq 0
    \end{cases}
\end{equation}

Eq.~\eqref{eq:bestfit_lnl} is important in the context of fitting, or finding the maximum-likelihood solutions of the model given the data, and as a selection criteria (see more on the latter below). While $\max_{ieo}\ln\overline{\mathcal{L}}_{ieo}$ is not strictly the maximum over the entire parameter space, under dense enough sampling it is an acceptable approximation. 

In addition to maximizing the likelihood, the inner products Eqs.~\eqref{eq:dh_ieo}--\eqref{eq:hh_ieo} evaluated at $d_L^0$ also allow one to marginalize over the distance parameter, as shown in \cite{Singer2016, roulet2024extrinsic}:
\begin{equation}
\label{eq:lnl_dist_marginalized}
    \ln\overline{\mathcal{L}}_{ieo} = \int \rmd d_L \pi(d_L) \exp\left( \frac{\langle d\mid h \rangle_{ieo}}{d_L/d_L^0} - \frac{1}{2}\frac{\langle h \mid h \rangle_{ieo}}{\left(d_L/d_L^0\right)^2} \right),
\end{equation}
where $\pi(d_L)$ distance prior.
The distance marginalization in this work is done using the \texttt{cogwheel} distance marginalization routines \cite{roulet2024extrinsic}. It does not benefit from the matrix form of the Eqs.~\eqref{eq:dh_ieo} and \eqref{eq:hh_ieo}. It is therefore beneficial to remove $(i,e,o)$ combinations that have small contribution to $\mathcal{Z}$ or little agreement with the data. Thus, we evaluate Eq.~\eqref{eq:lnl_dist_marginalized} only for samples with 
\begin{equation}
\label{eq:coherent_ML_selection}
    \ln\mathcal{L}_{ieo}^{\rm ML} \geq \max_{ieo}\ln\mathcal{L}^{\rm ML}_{ieo} - \Delta\ln\mathcal{L}^{\rm ML}
\end{equation}
with $\Delta \ln \mathcal{L}^{\rm ML}$ taken as the conservative value of 20 in this work.

Given $\ln\overline{\mathcal{L}}_{ieo}$ and points drawn directly from the factorizable prior $\pi({\bm \theta})=\pi(\thetaint)\pi(\thetaext)\pi(\phiref)\pi(d_L)$, the Bayesian evidence ratio $\mathcal{Z}$ can be evaluated as the ensemble average of the distance-marginalized likelihood:
\begin{equation}
    \mathcal{Z}=\int \rmd {\bm \theta}\mathcal{L}(\bm\theta)\pi(\bm \theta) \approx \frac{1}{N_{\rm int.}N_{\rm ext}N_{\phi}}\sum_{ieo}\overline{\mathcal{L}}_{ieo}
\end{equation}
As discussed in Section~\ref{subsec:integration_methods}, drawing points directly from the assumed physical prior is inefficient. Instead we draw samples from an informed prior $\pi'$, and weight them by the prior ratio $w$. We will elaborate on how this is done in the following section. The evidence is then evaluated as Eq.~\eqref{eq:Z_sum}, repeated here: 
\begin{equation*}
    \mathcal{Z}\approx \frac{1}{N_{\rm int.}N_{\rm ext}N_{\phi}}\sum_{ieo} w_i w_e w_o\overline{\mathcal{L}}_{ieo} 
\end{equation*}
with
\begin{equation}
    w_i=\frac{\pi\left(\thetaint^{(i)}\right)}{\pi'\left(\thetaint^{(i)}\right)},\,
    w_e=\frac{\pi\left(\thetaext^{(e)}\right)}{\pi'\left(\thetaext^{(e)}\right)},\,w_o=\frac{\pi\left(\phiref^{(o)}\right)}{\pi'\left(\phiref^{(o)}\right)}.
\end{equation}

$w_i$ and $w_e$ will be described in Sections~\ref{subsec:intrinsic_samples} and \ref{subsec:extrinsic_samples}. $\phiref^{(o)}$ are taken on a regular grid, making $w_o=1$.

\subsection{Intrinsic waveform banks}
\label{subsec:intrinsic_samples}
The samples $\big\{\thetaint^{(i)} \big\}_{i=1}^{N_{\rm int.}}$ used here are drawn ahead of time and stored in a bank.
This allows us to generate the intrinsic waveforms and store them, thereby replacing on-the-fly waveform generation with loading precomputed arrays.
This is beneficial when using more accurate, though computationally slower, waveform models. In particular, we use \texttt{IMRPhenomXODE}~\cite{Yu2023}, which incorporates higher-modes and spin-precession effects.
All waveforms are stored in relative-binning frequency resolution of $N_f=378$ points, which provide $\ln\mathcal{L}$ accuracy better than 1\%.

\begin{figure}
    \centering
    \includegraphics[width=\linewidth]{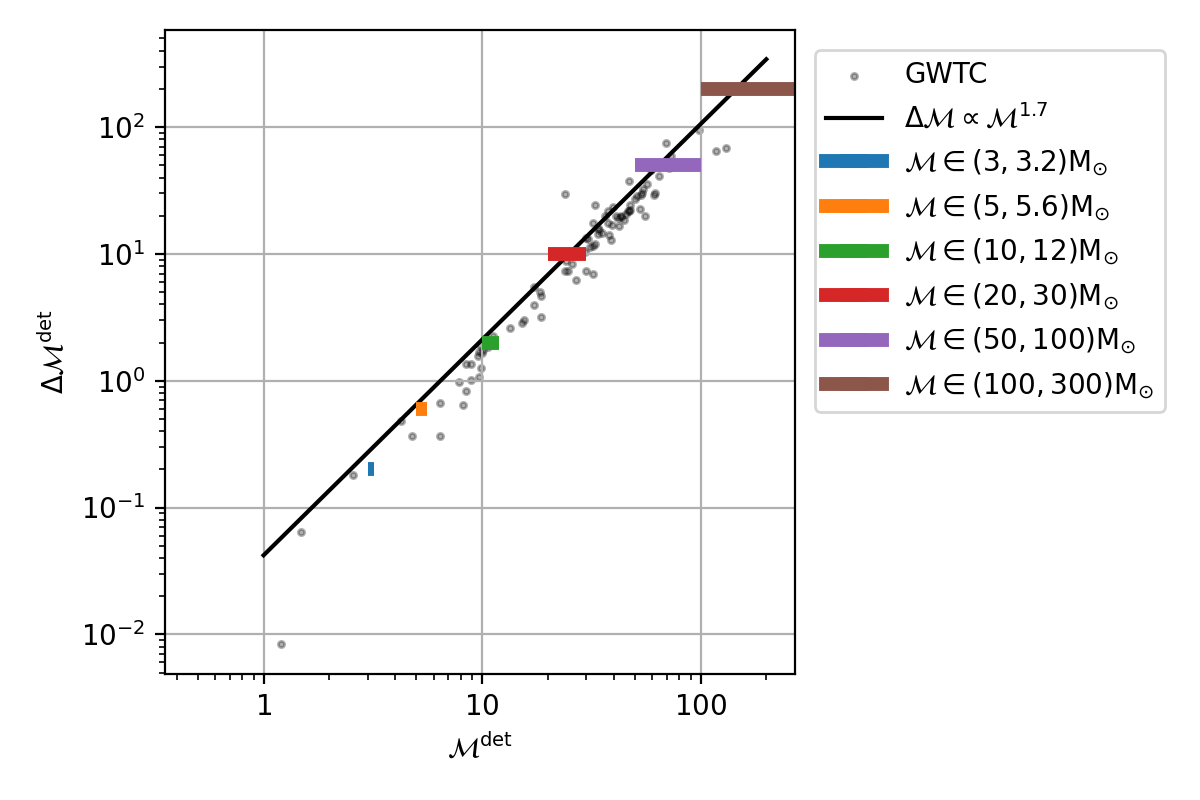}
    \caption[]{Empirical setting of the intrinsic waveform banks detector frame chirp mass ranges. Gray dots are the reported chirp mass and 90\% confident interval (CI) of 93 GWTC events (combination of GWTC-1, GWTC-2 and GWTC-3 confident events), scaled to SNR 8 by multiplying the width of the CI by ${\rm SNR/8}$. Solid black line is the empirical scaling we roughly used for the banks. The six banks used for tested in this paper are shown in horizontal solid lines, spanning the minimal to maximal $\mchirp$ in the x-axis, and also written in the legend.The width of the span in the y-axis.}
    \label{fig:bank_masses}
\end{figure}

Different banks are created for different mass ranges, using the heuristic that the chirp mass ranges scale as a power-law of the central chirp mass.
The prefactor and power-law are calibrated by-eye using the posterior samples of the 93 GWTC-3 events~\cite{Abbott2023_GWTC3}, as presented in Fig.~\ref{fig:bank_masses}. The mass ranges are scaled to SNR values of 8, typical of signals at the edge of detectability, by multiplying the confidence interval (CI) width by ${\rm SNR/8}$, following the reasoning of Fisher information matrix (FIM) analysis~\cite{Vallisneri2008PhysRevD}.

The physical $\pi(\thetaint)$ is taken to be uniform in $m_1, m_2$, 
limited to $\mchirp \in (\mchirp_{\rm min}, \mchirp_{\rm max})$ and $q \in (q_{\rm min}, 1)$. $q_{\rm min}$ is taken as $1/5$ in this work, which is less extreme than the validity limit of modern waveform generation codes, e.g.~\cite{Pratten2020PhRvD}. An isotropic binary orientation is assumed, implying $\cos\thetajn \sim \mathcal{U}(-1,1)$ and $\phijlhat \sim \mathcal{U}(0,2\pi)$~\cite{Cutler1994PhRvD,Roulet2022}.
The effective spin $\chieff$ and aligned-spin-difference coordinate $\cumchidiff$ are distributed as
\begin{align}
    \chieff &= \frac{\chi_{1,z}+q\chi_{2,z}}{1+q} \sim \mathcal{U}(-1,+1), \\ 
    \cumchidiff &= \frac{\chi_{1z}-\chi_{1z}^{\rm min}}{\chi_{1z}^{\rm max}-\chi_{1z}^{\rm min}} \sim \mathcal{U}(0,1).
\end{align}
The in-plane spin component $\chi_{1x}$ is sampled uniformly within the disc (conditioned on $\chi_{1z}, \chi_{2z}$).

The sampling distribution $\pi'(\thetaint)$ used to draw the waveform bank differs from the assumed physical prior.
Lower masses perform more cycles in band, so sampling is done uniformly in $\ln\mchirp$.
Lower mass ratios exhibit stronger higher-order modes, so sampling is done uniformly in $\ln q$.
$\thetajn$ is sampled uniformly between $0$ and $\pi$, and the spin priors remain unchanged. The probability of each intrinsic parameters sample is reweighted by
\begin{equation}
\label{eq:int_weights}
    w_i = \frac{\pi\left(\thetaint^{(i)}\right)}{\pi'\left(\thetaint^{(i)}\right)}
\end{equation}
to correct the over- or under-densities of samples relative to the target prior.

\subsection{Intrinsic Waveforms Pre-selection}
\label{subsec:preselection}
The runtime of the likelihood evaluations described in Section~\ref{subsec:lnlike_as_matmul} 
scales with the number of intrinsic samples, $N_{\rm int}$. As will be discussed in 
Section~\ref{subsec:extrinsic_samples}, extrinsic samples are drawn conditioned on selected 
intrinsic samples. Therefore, eliminating clearly incompatible intrinsic samples serves 
two purposes: first, it reduces the computational cost by limiting the number of intrinsic 
samples processed; second, it improves efficiency in the importance-sampling sense, as 
drawing extrinsic samples based on inconsistent intrinsic samples is wasteful. To identify 
and discard such intrinsic samples, we approximate the maximal coherent likelihood they 
could achieve using the incoherent sum of maximal likelihoods across detectors 
(see Appendix~\ref{app:single_detector_likelihood}). 

The selection criterion on the incoherently maximized likelihood is
\begin{equation}
\begin{aligned}
    \ln\mathcal{L}^{\rm inco.\,ML}_{i} &= \sum_k \max_{o,t}\ln\mathcal{L}^{\rm ML}_{ikot} \\
    &\geq \max_{j}\ln\mathcal{L}^{\rm inco.\,ML}_{j} - \Delta\mathcal{L}^{\rm inco.\,ML},
\end{aligned}
\end{equation}
with $\Delta\mathcal{L}^{\rm inco.\,ML}=20$ in this work. 
The number of intrinsic samples retained after the procedure is denoted $N_{\rm int.}'$. 
In its current implementation, the runtime of \texttt{dot-PE} is dominated by this 
pre-selection stage (see Appendix~\ref{app:complexity}).
\subsection{Extrinsic samples}
\label{subsec:extrinsic_samples}
Given data from an event, the phase space of possible extrinsic parameters is much larger than what the data allows, due to the small correlation timescale of the signals, and the geometric relation between sky-positions and signal arrival time at the individual detectors (Eq.~\eqref{eq:t_at_detector}). Therefore we adopt the adaptive-importance sampling method presented in \cite{roulet2024extrinsic} (described briefly in Appendix~\ref{app:ext_marginalization}) to draw extrinsic samples in an informed way. Given intrinsic parameters and the data, this method produces samples of $t_\oplus$ and $\los$ based on estimates of the times of arrival of the signal at each detector. $\psi$ is drawn uniformly from $(0,\pi)$. The method also allows us to evaluate the marginalized likelihood $\overline{\mathcal{L}}(\thetaint)$ per choice of intrinsic parameters, through a weighted ensemble average over $N_{\rm ext. marg}$ extrinsic samples, trapezoidal quadrature over $\phiref$, and analytic integration over $d_L$. For our purposes, we want to draw a single set of extrinsic samples, decoupled from intrinsic samples. This is done by aggregating extrinsic samples drawn using $N_c=16$ different intrinsic samples. This aggregation is motivated by previous works that have shown correlations between the inclination angle (treated here as intrinsic) and sky location \cite{Singer2014, Roulet2022}. After constructing the aggregated ensembles, the proposal distributions $\pi'$ are combined (as in Eq.~(30) of \cite{roulet2024extrinsic}), and importance-sampling weights are assigned to each extrinsic sample as 
\begin{equation} 
w_e=\frac{\pi\left(\thetaext^{(e)}\right)}{\pi'\left(\thetaext^{(e)} \,\middle|\, d,\big\{\thetaint^{(c)}\big\}_{c=1}^{N_c}\right)}.
\end{equation} 
We then draw $N_{\rm ext.}$ extrinsic samples uniformly from the combined ensemble and retain their importance-sampling weights. To ensure consistency with the data, we construct proposal distributions using only intrinsic samples that both passed the pre-selection process (Section~\ref{subsec:preselection}) and have $\ln\overline{\mathcal{L}}\geq0$. We further require two conditions on each proposal distribution constructed from a single intrinsic sample, that are motivated by the eventual effective sample size of the integral (marginalizing over $\phiref$):
\begin{equation}
    N_{\rm eff.} = 
    \frac{\left(\sum_{ie} w_e w_i \overline{\mathcal{L}}_{ie}\right)^2}{\sum_{ie}w_e^2 w_i^2 \overline{\mathcal{L}}_{ie}^2},
\end{equation}
where $\overline{\mathcal{L}}_{ie}$ is the distance and phase marginalized likelihood, for a given intrinsic-extrinsic sample pair. Under one simplistic limiting case, the likelihood per extrinsic sample is the same for all intrinsic samples used in the integral $\overline{\mathcal{L}}_{ie}=\overline{\mathcal{L}}_{e}$. The effective sample size can be factorized 
\begin{equation}
    N_{\rm eff.} = \frac{\left(\sum_i w_i\right)^2}{\sum_i w_i^2} \cdot 
    \frac{\left(\sum_{e}w_e \overline{\mathcal{L}}_{e}\right)^2}{\sum_{e} w_e^2 \overline{\mathcal{L}}_{e}^2},
\end{equation}
and the right-most ratio is the same in form as the effective sample size evaluated per proposal in \cite{roulet2024extrinsic}. We therefore require that each proposal (conditioned on $\thetaint$) satisfies $N_{\rm eff.}\geq 100$. In the opposite limiting case, the rows of $\overline{\mathcal{L}}_{ie}$ are uncorrelated, and we can replace $\overline{\mathcal{L}}_{ie}$ and $\overline{\mathcal{L}}_{ie}^2$ with their expectation values (using $\mathbb{E}$ to differentiate from the distance and phase marginalized value)
\begin{equation}
    N_{\rm eff.} = \frac{\left(\sum_e w_e\right)^2}{\sum_e w_e^2}\cdot \frac{\left(\sum_i w_i\right)^2}{\sum_i w_i^2}\cdot \frac{\mathbb{E}\left[\overline{\mathcal{L}}\right]}{\mathbb{E}\left[\overline{\mathcal{L}}^2\right]}.
\end{equation}
Under this assumption, the effective sample size is proportional to the prior-effective sample size 
\begin{equation}
    N_{\rm eff. prior} = \frac{\left(\sum_e w_e\right)^2}{\sum_e w_e^2}.
\end{equation}
We therefore require that each proposal satisfies $N_{\rm eff. prior}\geq 50$. Ensembles that do not satisfy both these conditions are discarded, and replaced by a new ensemble constructed from a different intrinsic sample.

\section{Performance}
\label{sec:performance}
We test the performance of our method against injections into synthetic noise realizations, using the Hanford--Livingston--Virgo detector network with O3 sensitivities. 1,024 injections are performed for each of 6 chirp mass ranges: $(3,3.2){\rm M}_\odot$, $(5,5.6){\rm M}_\odot$, $(10,12){\rm M}_\odot$, $(20,30){\rm M}_\odot$, $(50,100){\rm M}_\odot$, and $(100,300){\rm M}_\odot$. The injection prior are the same as our assumed physical priors, as used in the intrinsic bank creation and in extrinsic sample drawing: uniform $\mchirp$ in range, uniform $\ln q$ restricted to $q\in(0.2,1)$, uniform $\chieff\in(-1,1)$, uniform $C^{\perp}$ and isotropic in-plane spin for each binary component, isotropic sky position, uniform polarization angle and uniform luminosity--volume prior ($\pi(d_L)\propto d_L^2$). See \cite{Roulet2022} for definitions of the parameters. We impose on the injections $\langle h\mid h \rangle\in(70,200)$ appropriate for most detections. We also perform a small number of individual injections, to validate the performance of \texttt{dot-PE} for parameter estimation against nested sampling using \texttt{cogwheel}, see Section~\ref{subsec:PE}.

\subsection{Integration}
\label{subsec:integration}
We set a target accuracy of $\pm1$ in evaluating $\ln\mathcal{Z}$ using our work. In the Gaussian-posterior limit, $\ln\mathcal{Z}\propto\rho^2/2 + \mathcal{O}(\ln\rho)$. Although the detection limit could depend on the mass, it is roughly set to $\rho=8$. The distance and $\rho^2$ are related through (following Eqs.~\eqref{eq:lnlike_ieo_arb_dL}--\eqref{eq:bestfit_lnl}):
\begin{equation}
    \frac{\rho^2}{2}\approx \frac{\vert\langle d \mid h\rangle\vert^2 }{\langle h \mid h\rangle}\propto \frac{1}{d_L^2}
\end{equation}
 Following the error propagation from $\Delta\ln\mathcal{Z}$ to the fractional sensitive volume loss, $\Delta d_L^3/d_L^3$, we get 
\begin{equation}
    \frac{\Delta d_L^3}{d_L^3}=\frac{3}{2}\frac{\Delta d_L^2}{d_L^2}=\frac{3}{2}\frac{\Delta(\rho^2/2)}{\rho^2/2}=3\frac{\Delta \ln\mathcal{Z}}{\rho^2}\approx 5\%
\end{equation}
for $\rho=8$ and $\Delta\ln\mathcal{Z}=1$. This is comparable to the losses associated with template placement and PSD estimation within the IAS pipeline \cite{Venumadhav2019}.
We test the performance of the code in estimating $\mathcal{Z}$ for the injections specified in Section~\ref{sec:performance}. For each event in the set, we perform a run and evaluate $\ln\mathcal{Z}$. We either fix $N_{\rm int.}$ to the full banks size ($2^{16}$) and vary $N_{\rm ext.}$, or fix $N_{\rm ext.}$ to $2^{10}$ and vary $N_{\rm int.}$. $N_{\phi}=32$ in all runs. If $N_{\rm int.}$ is less than the whole bank size, we select a random starting point within the bank, and use all $N_{\rm int.}$ samples in a sequence, to exploit the low-discrepancy properties of the Halton sequence. The results are displayed in Fig.~\ref{fig:convergence}. The results demonstrate that evaluating the log of the evidence-ratio integral with accuracy of $\sim 1$ is possible, by working with $N_{\rm int.}$ of a few thousands and $N_{\rm ext.}$ of a few hundreds. It is also clear that $N_{\rm int.}'$, the number of samples not discarded by the pre-selection process, is a good proxy for the integration accuracy. This means that the method could iteratively apply the incoherent likelihood selection on smaller portions of the bank, until some target $N_{\rm int.}'$ is reached, and only then perform the coherent likelihood evaluations.

In post-analysis, it is possible evaluate the effective sample size (Eq.~\eqref{eq:n_effective}), using the posterior weights: 
\begin{equation}
\label{eq:posterior_element}
    N_{\rm eff.} = \frac{\left(\sum_{ieo}p_{ieo}\right)^2}{\sum_{ieo}p_{ieo}^2},\quad p_{ieo}=w_iw_ew_o\mathcal{L}_{ieo}.
\end{equation}

\begin{figure*}
    \centering
    \includegraphics[width=\linewidth]{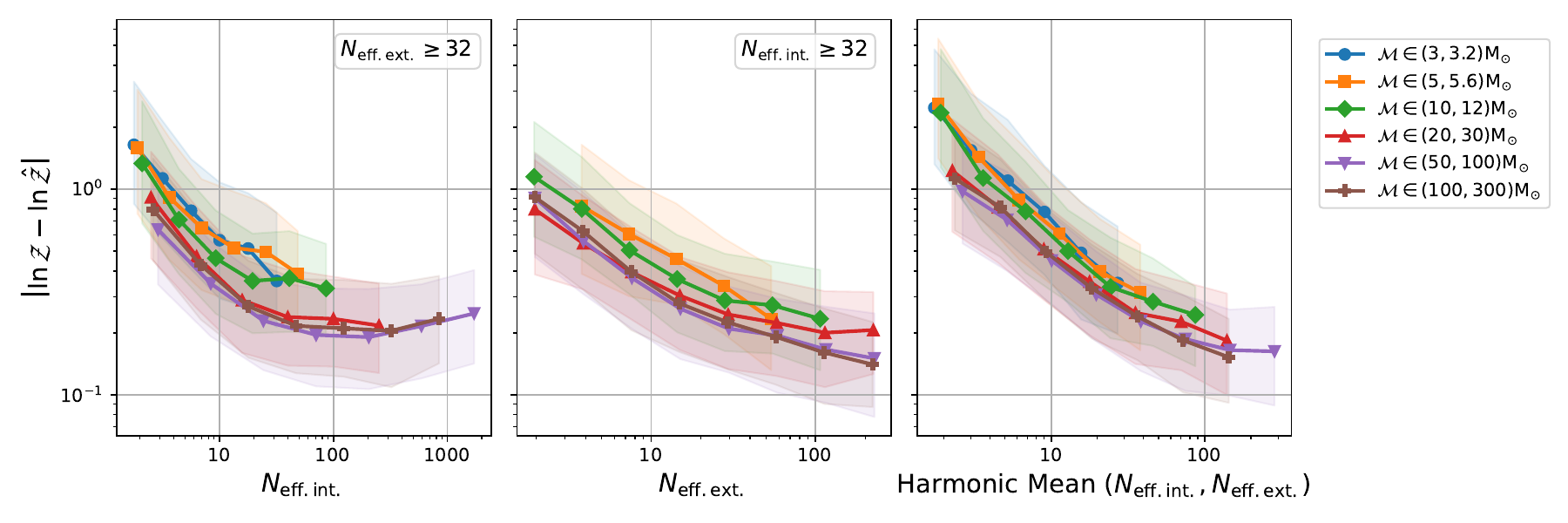}
\caption[]{Dependence of $\ln\mathcal{Z}$ error on effective sample sizes.
\textbf{Left:} Intrinsic ($N_{\rm eff. int.}$, \eqref{eq:int_ess}).
\textbf{Center:} Extrinsic ($N_{\rm eff. ext.}$, \eqref{eq:ext_ess}).
\textbf{Right:} Harmonic mean of the two.
Solid lines show medians; shaded bands span 25th–75th percentiles. See Section~\ref{subsec:integration} for details.
}
\label{fig:ess}
\end{figure*}

Similarly, we can marginalize over the orbital phases and the extrinsic (intrinsic) samples to get the effective samples size of intrinsic (extrinsic) samples:
\begin{align}
    \label{eq:int_ess}
    N_{\rm eff.\, int.} &= \frac{\left(\sum_{ieo}p_{ieo}\right)^2}{\sum_{i}\left(\sum_{eo}p_{ieo}\right)^2},\\
    N_{\rm eff.\, ext.} &= 
    \label{eq:ext_ess}
    \frac{\left(\sum_{ieo}p_{ieo}\right)^2}{\sum_{e}\left(\sum_{io}p_{ieo}\right)^2}.
\end{align}
A low effective sample size suggests that the distribution is not well represented by the weighted samples, and that the integral could be inaccurate. We show this by plotting $N_{\rm eff.\, int.}$ and $N_{\rm eff.\, ext.}$ against the error in $\ln\mathcal{Z}$ against the effective sample size, for all events used in Fig.~\ref{fig:convergence}. The results, shown in Fig.~\ref{fig:ess}, suggest that a good validity check for an integration is that the samples satisfy their harmonic mean $2(N_{\rm eff. int}^{-1}+N_{\rm eff. ext}^{-1})^{-1}\gtrsim10$. 

It is worth stressing the relation and differences between two similar sample sizes: $N_{\rm int.}'$, available after the incoherent pre-selection process but before the full coherent likelihood evaluation, and $N_{\rm eff.\,int.}$, available only after the run is complete. As can be seen in Figs.~\ref{fig:convergence} and \ref{fig:ess}, they both serve as indicators to the integration accuracy, $N_{\rm eff. int.}$ being the more predictive of the two. $N_{\rm int.}'$ can be thought as a rough estimate of the effective intrinsic sample size, that treat all accepted samples as equiprobable.

\begin{figure}
    \centering
    \includegraphics[width=\linewidth]{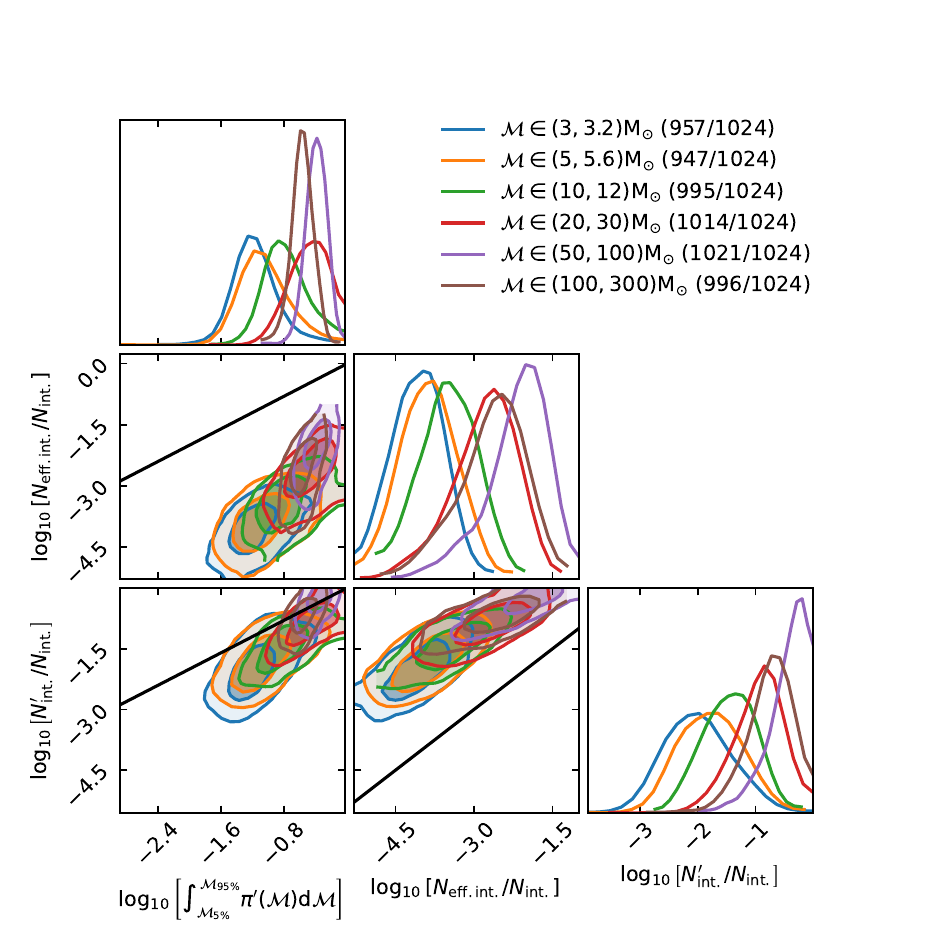}
    \caption[]{The relation between $N_{\rm int.}'$ (after incoherent likelihood selection), effective intrinsic samples size $N_{\rm eff.\,int.}$ and $\mchirp$ confidence--interval prior integral (Eq.~\eqref{eq:mchirp_CI_integral}). The results for each bank are based on a number of runs with at $N_{\rm eff.}\geq20$ shown in the legend, out of a total 1,024. 1:1 relations are plotted in a solid black lines.}
    \label{fig:selection_and_ESS}
\end{figure}

The relevant confidence intervals of $\mchirp$ are determined from the same injection runs analyzed throughout this section. $N_{\rm int.}'$ is then estimated by counting the number of samples within the confidence interval,
\begin{equation}
\label{eq:mchirp_CI_integral}
    N_{\rm int.}'\approx N_{\rm int.}\cdot \int_{\mchirp\in {\rm CI}} \pi'(\mchirp){\rmd}\mchirp
\end{equation}
with an accuracy of roughly an order of magnitude, due to effects of precession and higher modes. In Fig.~\ref{fig:selection_and_ESS} we show the relation between $N_{\rm int.}'$, $N_{\rm eff.\, int.}$ and Eq.~\eqref{eq:mchirp_CI_integral} for these injections, using denser banks of $N_{\rm int.}=2^{18}$ waveforms. Since we include the mass confidence interval, we use only runs with at least 20 samples (or $N_{\rm eff.}\geq40$).

With this, we can conclude by summarizing how to create a bank or banks: we start with the target $\ln\mathcal{Z}$ accuracy we aim for. From it, we estimate the required $N_{\rm int.}'$ using the result as shown in Fig.~\ref{fig:convergence}. We then select $\mchirp$ bounds for the bank, balancing the consideration that the wider the banks are, the less of them are needed, against the fact that a smaller fraction of samples will be in the CI-prior integral. The integral itself can be estimated from the posterior samples of similar events, from an injection study, or from a Fisher information matrix analysis (but consider the limitation of FIM anaylsis in low SNR and non-Gaussian posteriors~\cite{Vallisneri2008PhysRevD}). In addition, the $\mchirp$ ranges should be no smaller than the support of typical CI in the expected SNR, estimated in Fig.~\ref{fig:bank_masses}. Given the integral, and considering a typical acceptance rates going as low as 0.1 of the integral, based on Fig.~\ref{fig:selection_and_ESS}, the total bank size can be taken by inverting the relation in Eq.~\eqref{eq:mchirp_CI_integral}.

\subsection{Parameter Estimation}
\label{subsec:PE}
Using the approximation that the effective sample size is roughly the number of individual samples in the distribution, we limit the number of samples $N_s$ drawn by \texttt{dot-PE} to $N_{\rm eff.}/2$.

To demonstrate the performance of the method in sampling a posterior distribution, we performed PE runs on 3 synthetic events in different mass ranges. We compare the performance of our method against \texttt{cogwheel} \cite{Roulet2022}, using marginalization over extrinsic parameters \cite{roulet2024extrinsic} and the \texttt{\texttt{nautilus}} sampler \cite{Lange2023}. The injection parameters are presented in Table \ref{tab:single_events}. The PE using the method presented in this work is done using $N_{\rm int.}=2^{16}$, $N_{\rm ext.}=2^{11}$, $N_\phi=32$. The \texttt{cogwheel} nested sampler runs are performed with 1,000 live-points. We limit \texttt{cogwheel}'s $\mchirp$ and $q$ to the same ranges as the used intrinsic sample banks. The results are shown in Figs.~\ref{fig:event_1}--\ref{fig:event_3} and demonstrate that the two methods are in close agreement.

\begin{figure*}
    \centering
    \includegraphics[width=\linewidth]{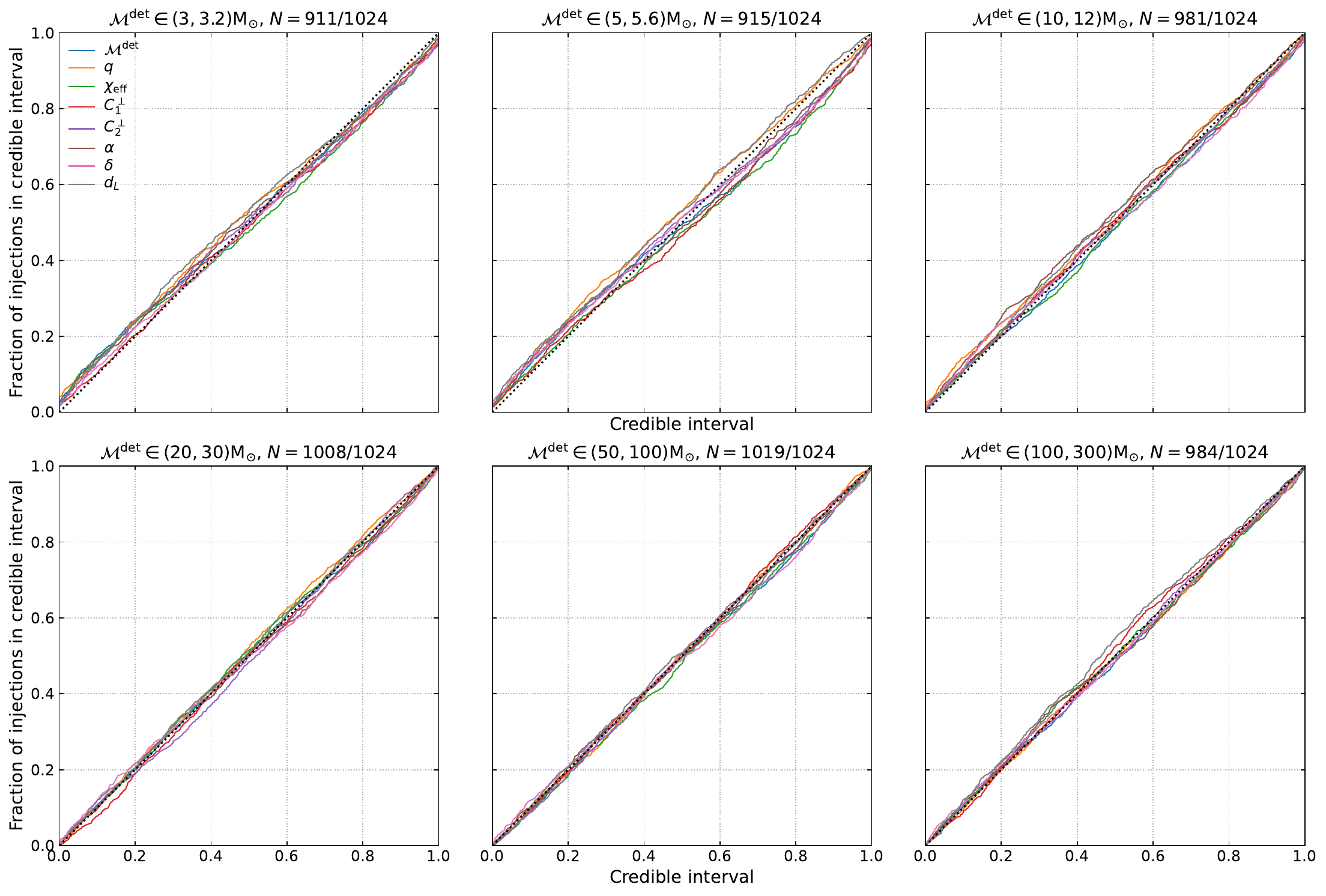}
    \caption{Probability--probability (P--P) plots at six mass ranges. Different colored lines stand for different well-measured parameters, which are defined in \cite{Roulet2022}. The fraction in each title stands for the number of runs that produced more than 30 samples with $\langle h\mid h\rangle\in(70,200)$, as imposed on the injections themselves, out of all runs performed. The colored lines agree with the diagonal 1:1 lines (black dotted), meaning that the samples from the posterior are drawn correctly.}
    \label{fig:pp_plots}
\end{figure*}

To evaluate the performance and bias of parameter estimation runs, we use \texttt{dot-PE} to sample the posteriors of the synthetic injections describe in the beginning of this section. We impose the same $\langle h \mid h \rangle\in(70,200)$ criterion on the posterior samples used to generate the plots. The parameter estimation runs are performed with intrinsic samples banks of $N_{\rm int.}=2^{16}$, $N_{\rm ext.}=2^{10}$ extrinsic samples and $N_{\phi}=64$. A few percent of the events resulted in less than 30 samples (overall, or in the allowed $\langle h\mid h \rangle$ range) and are excluded from the analysis. Per injection and set of samples, we evaluate the cumulative density functions (CDF) of several well-measured parameters. The evaluated CDF values are expected to distribute uniformly in $(0,1)$. The resulting probability--probability (P--P) plots are shown in Fig.~\ref{fig:pp_plots}, where agreement between the empirical CDFs (solid colored lines) and the ideal diagonal (black dotted lines) indicates the accuracy of the method.

\section{Conclusions}
\label{sec:conclusions}
We have presented a novel approach to the posterior estimation problem of CBC GW events. We split the 15 dimensions of parameter space into 4 subsets, according to their functional form in the likelihood function. This allows us to evaluate the (distance-marginalized) likelihood for each intrinsic-extrinsic-phase combination, using a sequences of matrix multiplications. We implemented this method in \texttt{dot-PE}, a pure Python package. The method allows us to integrate the Bayesian evidence ratio, to accuracy of $\pm1$ in $\ln\mathcal{Z}$ within a few minutes (on a single CPU core). The runtime can be greatly reduced by trivial parallelization (already implemented in the current code) and GPU usage. The relations between the number of samples used, $N_{\rm int.}$, $N_{\rm ext.}$ and $N_{\phi}$ to the runtime and performance of the integration or parameter estimation is predictable and well understood, as demonstrated in Section~\ref{sec:performance} and Appendix~\ref{app:complexity}. For the purpose of a search, \texttt{dot-PE} could produce the evidence ratio, which is the optimal test statistic and includes the effects of precession and higher modes, for many thousands of candidates provided by a traditional match-filtering search. 

\section*{Acknowledgments}

BZ and JM are supported by the Israel Science Foundation, NSF-BSF. BZ is supported also by a research grant from the Willner Family Leadership Institute for the Weizmann Institute of Science.
JR acknowledges support from the Sherman Fairchild Foundation.
TV acknowledges support from NSF grants 2012086 and 2309360, the Alfred P. Sloan Foundation through grant number FG-2023-20470, the BSF through award number 2022136, and the Hellman Family Faculty Fellowship.

This research has made use of data, software and/or web tools obtained from the Gravitational Wave Open Science Center (\url{https://www.gw-openscience.org/}), a service of LIGO Laboratory, the LIGO Scientific Collaboration and the Virgo Collaboration. LIGO Laboratory and Advanced LIGO are funded by the United States National Science Foundation (NSF) as well as the Science and Technology Facilities Council (STFC) of the United Kingdom, the Max-Planck-Society (MPS), and the State of Niedersachsen/Germany for support of the construction of Advanced LIGO and construction and operation of the GEO600 detector. Additional support for Advanced LIGO was provided by the Australian Research Council. Virgo is funded, through the European Gravitational Observatory (EGO), by the French Centre National de Recherche Scientifique (CNRS), the Italian Istituto Nazionale di Fisica Nucleare (INFN) and the Dutch Nikhef, with contributions by institutions from Belgium, Germany, Greece, Hungary, Ireland, Japan, Monaco, Poland, Portugal, Spain.

\appendix
\section{Relative Binning}
\label{app:relative_binning}
Relative binning, or heterodyning, is a method to reduce the frequency resolution of waveforms required to perform likelihood evaluations \cite{Cornish2013, Zackay2018, Leslie2021}.
We follow the formulation of \cite{Roulet2024}, implemented in \texttt{cogwheel} and reproduced here for completeness and specific implementation details. The core of the idea is that for two frequency domain waveforms $h$ and $h'$ not too far apart in parameter space, their ratio $h'(f) / h(f)$ is a slowly varying function of $f$. This statement remains true per mode and polarization. The slow frequency variation statement holds for the integrands of the likelihood integrals per detector, polarization and mode $\langle d_k \mid h_{kmp}\rangle$ and $\langle h_{kmp}\mid h_{km'p'}\rangle$. Given some reference waveform $h^0$, the relative binning weights $W^{\langle d \mid h \rangle}_{kmpb}$ and $W^{\langle h \mid h \rangle}_{kmm'pp'b}$ are found as 
\begin{equation}
\label{eq:dh_weights}
    W^{\langle d \mid h \rangle}_{kmpb} = \frac{\langle d_k \mid h^0_{mp} s_b \rangle }{(h_{mp}^0(f_b))^\ast} 
\end{equation}
where $b$ are the indices of the spline points on the frequency grids $f_b$ and of the spline functions $s_b(f)$, that satisfy $s_b(f_b')=\delta_{bb'}$. Similarly,
\begin{equation}
\label{eq:hh_weights}
W^{\langle h \mid h \rangle}_{kmm'pp'b} = \frac{\langle h^0_{mp} \mid h^0_{m'p'} s_b \rangle}{h^0_{mp}(f_b) (h^0_{m'p'}(f_b))^\ast}
\end{equation}
In both \eqref{eq:dh_weights}, \eqref{eq:hh_weights} no summation over $k$ is assumed. The inner products  given $\thetaint$, $\thetaext, \phiref$ and $D$ are then:
\begin{align}
\label{eq:d_h}
    \langle d \mid h\rangle &\approx \sum_{k,m,p,b} W^{\langle d \mid h \rangle}_{kmpb}h_k(f_b)^\ast\\
    \label{eq:h_h}
    \langle h \mid h \rangle &\approx \sum_{kmm'pp'b} W^{\langle h\mid h\rangle}_{kmm'pp'b}h_k(f_b) h_k^\ast(f_b)
\end{align}
In equation \eqref{eq:dh_weights}, \eqref{eq:hh_weights}, any selection of $\thetaext$ for $h^0$ will cancel out due to denominators, with the sole exception of the time-shift exponents. Still, for small enough time-shifts, $\exp(2i \pi f \Delta t)$ will be smooth and the weights still provide the correct result. A typical size for the frequency axis $f$ is $\sim 10^5$, while the number of bins $b$ sufficient for $\sim1\%$ accuracy is $\sim 10^2$.

By selecting fixed frequency bins $f_b$, the intrinsic-waveform banks can be computed and saved only on them ($h_{impb}$), and not in full frequency resolution ($h_{impf})$.

\section{Optimizing likelihood over extrinsic parameters with single-detector data} 
\label{app:single_detector_likelihood}
The derivation in the appendix is not novel, and has been applied by several authors, when attempting to perform a search with precessing templates \cite{2023PhRvD.108l3016M, Pan2015PhysRvD, 2016PhRvD..94b4012H}. We present it here for completeness.
For single-detector data, the sky-position becomes degenerate with overall amplitude, polarization angle and time, in such a manner that the only contribution is from the prior. This is true provided the signal does not originate from a detector null ($\mathcal{R}_k\neq0$). Any change in time-delay from the center of the Earth, due to a change in sky-position, can be counteracted by a different choice of $t_\oplus$ (Eq.~\eqref{eq:t_at_detector}). Any change in the detector response amplitude $\mathcal{R}_k$ (Eqs.~\eqref{eq:response_plus}-\eqref{eq:response_cross}) can be compensated by changing the distance (Eq.~\eqref{eq:distance_scaling}):
\begin{align}
(\lambda, \varphi) &\rightarrow (\lambda', \varphi'),\\
t_\oplus &\rightarrow t_\oplus' = t_\oplus -  \frac{{\bm r}_k \cdot \hat{\bm n}(\lambda, \varphi)}{c} +  \frac{{\bm r}_k \cdot \hat{\bm n}(\lambda', \varphi')}{c},\\
d_L&\rightarrow d_L'=d_L\frac{\mathcal{R}_k(\lambda',\varphi')}{\mathcal{R}_k(\lambda, \varphi)}.
\end{align}
The implication of this degeneracy is that the best-fit amplitude factor, $x_{kp} = F_{kp}d_L^0/d_L$ can be solved for. We replace $\thetaext$ with a Cartesian product of $t$ and $\phiref$ on regular grids. For any choice of intrinsic parameters (indexed $i$), time (indexed $t$) and orbital phase (indexed $o$), the following components can be computed:
\begin{equation}
\label{eq:dh_itop}
    \langle d | h\rangle_{itop} = \Re \sum_{mb}\, W^{\langle d\vert h\rangle}_{kmpb} h_{impb}\, T_{tkb} e^{-im \phi_{\rm ref}^{(o)}},
\end{equation}
and similarly:
\begin{equation}
    \langle h | h\rangle_{iopp'} = \sum_{mm'b}\, W^{\langle h\vert h\rangle}_{kmm'pp'b} h_{impb}\, h^\ast_{im'p'b} e^{i(m-m') \phi_{\rm ref}^{(o)}},
\end{equation}
where the two expressions are real. The likelihood is then:
\begin{equation}
    \ln\mathcal{L}_{iot} = \sum_p  \langle d\vert h\rangle_{iotp} x_{kp}- \frac{1}{2} \sum_{pp'}\langle h \vert h \rangle_{iopp'} x_{kp}x_{kp'}
\end{equation}

For each selection of $(i,o,t)$, the maximum is found by optimizing a quadratic equation in two-variables $x_{k+}, x_{k\times}$:
\begin{equation}
\label{eq:x_ML}
    x_{iotkp}^{\rm ML} = \sum_{p'}\left[\langle h \vert h \rangle_{io} \right]^{-1}_{pp'}\langle d\vert h\rangle_{iotp'},
\end{equation}
where the first term is the $2\times2$ inverse matrix with specific $(i,o)$ selection. Plugging it back into the likelihood yields
\begin{equation}
\label{eq:single_det_ML}
    \ln\mathcal{L}_{iot}^{\rm ML} = \frac{1}{2} \sum_{pp'} \langle d \vert h\rangle_{iotp} \left[\langle h\vert h\rangle_{io}\right]^{-1} \langle d \vert h \rangle_{iotp'}
\end{equation}
Maximizing over $\phi_{\rm ref.}$ and $t_{\rm \oplus}$ on discrete grids provides the maximal likelihood per intrinsic sample. Summing the likelihoods independently maximized in different detectors provides the incoherent (across detectors) maximal-log-likelihood, $\ln\mathcal{L}^{\rm inco.\, ML}_{\rm i}$, which we use as a acceptance criteria for waveforms in the bank, prior to the drawing of extrinsic samples. 

\section{Extrinsic parameter marginalization}
\label{app:ext_marginalization}
For completeness, we summarize the method introduced in \cite{roulet2024extrinsic} for marginalization of the likelihood over extrinsic parameters. 
As a preparation step, a large number of sky-positions is drawn uniformly on the sky, using a QMC sequence, then binned into small equal-area tiles. Per sky position, the detector locations allow us to evaluate the arrival time $\tau$ difference between the different detectors (assuming, e.g., $t_\oplus=0$)
\begin{equation}
\boldsymbol{\tau}=(\tau_1,\tau_2,,\ldots) \quad \text{and so on for $N_{\rm det.}$ detectors}
\end{equation}
The arrival times are then binned in some finite time resolution. As can be seen in Eq.~\eqref{eq:t_at_detector}, these time delays between detectors define a constrained surface in sky-position space.

Given detector data and an intrinsic sample $\thetaint$, we can evaluate, for each detector $k$, the incoherent (across modes and polarizations) $\langle d \mid h\rangle_{kt}$ timeseries and $\langle h \mid h\rangle_k$ (constant), for a distance-optimized SNR-squared timeseries $\rho^2_k(t)$:
\begin{equation}
    \rho_k^2(t) = \frac{|\langle d \mid h\rangle_{kt}|^2}{2\langle h \mid h\rangle_k}.
\end{equation}
This timeseries can be converted to arrival time probability per detector:
\begin{equation}
    p_k(t) \propto \exp\left(\beta(\rho_k^2(t)-\max_t \rho^2_k(t))\right)
\end{equation}
Where $\beta\sim0.5$ is an inverse-temperature regularization term, pre-optimized to maximize the effective sample size. A large number $N_\mathrm{QMC}$ of arrival-time samples $\left\{\boldsymbol{\tau}_q\right\}_{q=1}^{N_\mathrm{QMC}}$ are drawn from the distributions (through their cumulatives). The exclusion of non-physical arrival times inconsistent with any sky position is discussed in detail in \cite{roulet2024extrinsic}, but omitted here. The samples of arrival time at each detector are then translated to a global arrival times $t_\oplus$ and sky-positions $\lambda,\varphi$, by drawing from all sky-position samples consistent (within the time resolution specified) with the inter-detector time delays. Since the intrinsic components of the waveforms are already known, it is relatively quick to evaluate the likelihood components (adopting the notations used in Section~\ref{subsec:lnlike_as_matmul}) $\langle d\mid h\rangle_{em}$ and $\langle h \mid h\rangle_{em}$ (for the given intrinsic samples and all extrinsic samples and modes), then apply different $\phiref$, marginalize over distance, and average over phase for $\overline{\mathcal{L}}_{\phi d_L}(\thetaext|\thetaint)$. The marginalized likelihood of the intrinsic sample can now be evaluated as 

\begin{equation}
    \overline{\mathcal{L}}(\thetaint)=\frac{1}{N_\mathrm{QMC}}\sum_{e=1}^{N_\mathrm{QMC}}\frac{\pi\left(\thetaext^{(e)}\right)}{\pi'\left(\thetaext^{(e)}\right)}\overline{\mathcal{L}}_{\phi d_L}(\thetaext^{(e)}|\thetaint)
\end{equation}
where the proposal distribution $\pi'$ (or $P$ in \cite{roulet2024extrinsic}) includes the time of arrivals and phase space consistent with it. The effective sample size of the $N_\mathrm{QMC}$ samples can be evaluated through the summands using Eq.~\eqref{eq:n_effective}. If the result is not satisfactory, a new, smoother arrival-time distribution is created by averaging the current distribution with a kernel-density estimation (on the current weighted samples) performed with a heavy-tailed Cauchy kernel. New arrival times are then drawn from the updated distribution. This can be repeated several times, making the proposals smoother in each iteration, until either some maximal number of iterations is reached or the effective sample size is satisfactory.

Other than evaluating the marginalized likelihood, the summands allow us to draw extrinsic samples for each intrinsic sample, and similarly, distance and phase can be drawn per $(\thetaint,\thetaext)$ pair. 

The three major modifications made in this work to this method are
\begin{enumerate}
    \item The effective sample size is evaluated both with and without the likelihood.
    \item The proposals $\pi'$ from $N_c=16$ different intrinsic samples are drawn, to decouple between the intrinsic and extrinsic samples (mainly due to inclination correlations with sky positions, \cite{Singer2014, Roulet2022}).
    \item The extrinsic samples are drawn uniformly, and the prior-ratio is taken as the importance-sampling weights $w_e$.
\end{enumerate}

\section{Complexity and runtime}
\label{app:complexity}
To test the runtime of the code in realistic settings, we chose $N_{\rm int.}$ and $N_{\rm ext.}$ per intrinsic sample bank based on the results of Section~\ref{subsec:integration}. In each of $N_{\rm int.}$ and $N_{\rm ext.}$, we selected the minimal value so that the median $\ln\mathcal{Z}$ accuracy will be $\pm1$. A single run was performed on the 1,024 injections per mass range described in Section~\ref{sec:performance}. The runs were performed on a single CPU core from the \texttt{Xeon Gold 6226R} CPU with clock rate of 2.10 GHz. As in Section~\ref{subsec:integration}, the errors are measured relative to the $\ln\mathcal{Z}$ computed using separate denser banks of size $N_{\rm int.}=2^{18}$, and with $N_{\rm ext.}=2^{10}$. The runtimes and typical dimensions are shown in Table~\ref{tab:working_point}. We also analyze the expected computational cost (in floating point operations, FLOPs) in Table~\ref{tab:flops}. We do not include \texttt{cogwheel} implemented operations, namely drawing extrinsic samples and evaluating the distance-marginalized likelihood. All operations not appearing in Table~\ref{tab:flops} are regarded as ``Overhead" in Table~\ref{tab:working_point}. 
The core \texttt{dot-PE} linear-algebra operations are dominated in runtime and FLOP count by the pre-selection process. The fractional costs of the pre-selection and overhead appear in Table~\ref{tab:working_point}, and they suggest the lower limits of runtime under improvements to either the overhead or to the linear-algebra functions (e.g., using GPUs or a large number of CPUs.). 

The number of distance-marginalizations $N_{\rm margin.}$ is typically within an order of magnitude of
$N_{\rm int.} \cdot N_{\rm ext.}$, given $N_{\phi} = 32$ phase values.
Since the log-likelihood exhibits approximately cosine-like dependence on $\phiref$,
only a few $\phiref$ values tend to pass the $\ln\mathcal{L}_{ieo}^{\rm ML}$ cut for each $(i,e)$ combination.
This indicates that the maximum-likelihood criterion effectively selects a small number of phases
per intrinsic–extrinsic pair, which in turn suggests that both the intrinsic pre-selection process
and the extrinsic sampling strategy in \texttt{dot-PE} are efficient.

\begin{table*}[ht]
\centering
\renewcommand{\arraystretch}{1.4}
\begin{tabular}{lcccccc}
\toprule
 &  $(3,3.2)\,{\rm M}_{\odot}$  & $(5,5.6)\,{\rm M}_{\odot}$  & $(10,12)\,{\rm M}_{\odot}$ & $(20,30)\,{\rm M}_{\odot}$ & $(50,100)\,{\rm M}_{\odot}$ & $(100,300)\,{\rm M}_{\odot}$ \\
\midrule 
$(N_{\rm int.},\, N_{\rm ext.})$  & ($2^{16}$, $2^{10}$) & ($2^{16}$, $2^{9}$) & ($2^{14}$, $2^{6}$) & ($2^{11}$, $2^{5}$) & ($2^{9}$, $2^{5}$) & ($2^{11}$, $2^{5}$) \\
$N_{\rm int.}'/N_{\rm int.}$  & $0.01^{0.03}_{0.005}$ & $0.02^{0.05}_{0.01}$ & $0.1^{0.1}_{0.02}$ & $0.2^{0.3}_{0.1}$ & $0.6^{0.8}_{0.4}$ & $0.3^{0.4}_{0.2}$ \\
$N_{\rm margin.}/\left(N_{\rm int.}'N_{\rm ext.}N_{\phi}\right)$  & $0.01^{0.01}_{0.003}$ & $0.01^{0.01}_{0.005}$ & $0.02^{0.03}_{0.01}$ & $0.03^{0.1}_{0.02}$ & $0.1^{0.1}_{0.03}$ & $0.2^{0.7}_{0.1}$ \\
$\left|\Delta \ln \mathcal{Z}\right|$
& $0.52^{1.07}_{0.20}$ & $0.45^{0.93}_{0.21}$ & $0.58^{1.12}_{0.28}$ & $0.59^{1.10}_{0.29}$ & $0.55^{1.07}_{0.23}$ & $0.68^{1.22}_{0.31}$ \\
Runtime (min.)  & $14.7^{15.2}_{12.9}$ & $14.8^{15.4}_{12.9}$ & $5.7^{6.0}_{4.9}$ & $2.8^{3.1}_{2.4}$ & $2.6^{2.8}_{2.1}$ & $3.0^{3.3}_{2.6}$ \\
Pre-selection/Total runtime  & $0.81^{0.82}_{0.79}$ & $0.81^{0.83}_{0.79}$ & $0.54^{0.56}_{0.52}$ & $0.14^{0.14}_{0.13}$ & $0.04^{0.04}_{0.04}$ & $0.13^{0.14}_{0.12}$ \\
Overhead/Total runtime  & $0.19^{0.21}_{0.18}$ & $0.18^{0.20}_{0.17}$ & $0.46^{0.48}_{0.44}$ & $0.86^{0.87}_{0.85}$ & $0.96^{0.96}_{0.96}$ & $0.87^{0.88}_{0.86}$ \\
\bottomrule
\end{tabular}
\caption{
Summary of runs aimed $\ln\mathcal{Z}$ accuracy of $\pm1$. $N_{\rm int.}$ and $N_{\rm ext.}$ are chosen based on the accuracies presented in Fig.~\ref{fig:convergence}. Subscripts and sup-scripts stand for 0.25 and 0.75 quantile, based on 1,024 injections. Runtimes quoted correspond to using a single CPU, see body of text.}
\label{tab:working_point}
\end{table*}

\begin{table*}[ht]
\centering
\renewcommand{\arraystretch}{1.2}
\begin{tabular}{llcccccc}
\toprule
\textbf{Operation} &  \textbf{FLOPs} &$(3,3.2)\,{\rm M}_{\odot}$  & $(5,5.6)\,{\rm M}_{\odot}$  & $(10,12)\,{\rm M}_{\odot}$ & $(20,30)\,{\rm M}_{\odot}$ & $(50,100)\,{\rm M}_{\odot}$ & $(100,300)\,{\rm M}_{\odot}$  \\
\midrule
Pre-selection & $6\,N_{\rm int.}\, N_{t}\, N_{f}\, N_{m}\, N_{\rm det}\, N_{p}$ & $2 \times 10^{11}$ & $2 \times 10^{11}$ & $6 \times 10^{10}$ & $8 \times 10^{9}$ & $2 \times 10^{9}$ & $8 \times 10^{9}$ \\
$\langle d \mid h \rangle_{iem}$ & $8\,N_{\rm int}'\, N_{\rm ext}\, N_f\, N_m\, N_{\rm det}\, N_p$ & $6 \times 10^{10}$ & $5 \times 10^{10}$ & $4 \times 10^{9}$ & $9 \times 10^{8}$ & $7 \times 10^{8}$ & $1 \times 10^{9}$ \\
$\langle h \mid h \rangle_{iem}$ & $8\,N_{\rm int}'\, N_{\rm ext}\, N_M\, N_{\rm det}\, N_p^2$ & $9 \times 10^{8}$ & $9 \times 10^{8}$ & $2 \times 10^{8}$ & $8 \times 10^{7}$ & $7 \times 10^{7}$ & $1 \times 10^{8}$ \\
$\langle d \mid h \rangle_{ieo}, \langle h \mid h \rangle_{ieo}$ & $8\,N_{\rm int}'\, N_{\rm ext}\, (N_m + N_M)\, N_{\phi}$ & $3 \times 10^{9}$ & $2 \times 10^{9}$ & $2 \times 10^{8}$ & $4 \times 10^{7}$ & $4 \times 10^{7}$ & $6 \times 10^{7}$ \\
$\ln\mathcal{L}^{\rm ML}_{ieo}$ & $4\,N_{\rm int}'\, N_{\rm ext}\, N_{\phi}$ & $1 \times 10^{8}$ & $9 \times 10^{7}$ & $7 \times 10^{6}$ & $2 \times 10^{6}$ & $1 \times 10^{6}$ & $2 \times 10^{6}$ \\
\bottomrule
\end{tabular}
\caption{
Summary of computational operations and estimated floating-point operation counts (FLOPs) in \texttt{dot-PE}. Not included here are \texttt{cogwheel} implemented routines to draw the extrinsic samples and distance-marginalize the likelihoods. In the runs, $N_{\rm int.}$ and $N_{\rm ext.}$ are taken from Table~\ref{tab:working_point}. $N_t=128$, $N_\phi=32$, $N_m=4$, $N_p=2$, $N_M=10$, $N_{\rm det}=3$, $N_f=378$. See more details in the body of the text.}
\label{tab:flops}
\end{table*}

\section{Parameter estimation examples}
\label{app:pe}
We compare the results of PE runs performed with our method against the \texttt{cogwheel} \cite{Roulet2022} sampling code, using extrinsic-parameter marginalization techniques \cite{roulet2024extrinsic}, and the \texttt{nautilus} nested sampler \cite{Lange2023}. The parameters of the injections are presented in Table~\ref{tab:single_events}.

\begin{table*}[t]
    \centering
    \caption{Parameters of example synthetic events in Figs.~\ref{fig:event_1}--\ref{fig:event_3}.}
    \begin{tabular}{lccccccccccccc}
        \toprule
         & $\mchirp$(${\rm M}_{\odot}$) & $q$ & ${\bm \chi}_1$ & ${\bm \chi}_2$ & $\alpha$ & $\delta$ & $\cos\iota$ & $\psi$ & $\phiref$ &$d_L$(Gpc)&Network & $\ln\mathcal{L}$ \\
        \midrule
        Event 1 & 24 & 2/3 & (0.5,0.5,0.6) & (0.5,0.5,0.6) & 1.44 & 0.89 & 0.05 & 4.41 & 0.36 & 1.5 & HLV & 43.3 \\
        Event 2 & 70  & 1/2  & (0.5,-0.5,0.6)  & (-0.5,0.5,0.6)  & 5.51  & -0.48 & -0.67  & 0.59  & 2.98 & 3.0 & HLV & 43.9 \\
        Event 3 & 200 & 1/2 & (0.3, 0.7, 0.6) & (0.9, 0.1, 0.1) & 2.34 & 0.7 & -0.71 & 2.30 & 2.57 & 4.0 & HLV & 29.9\\
        \bottomrule
    \end{tabular}
    \label{tab:single_events}
\end{table*}

\begin{figure*}
    \centering    \includegraphics[width=\linewidth]{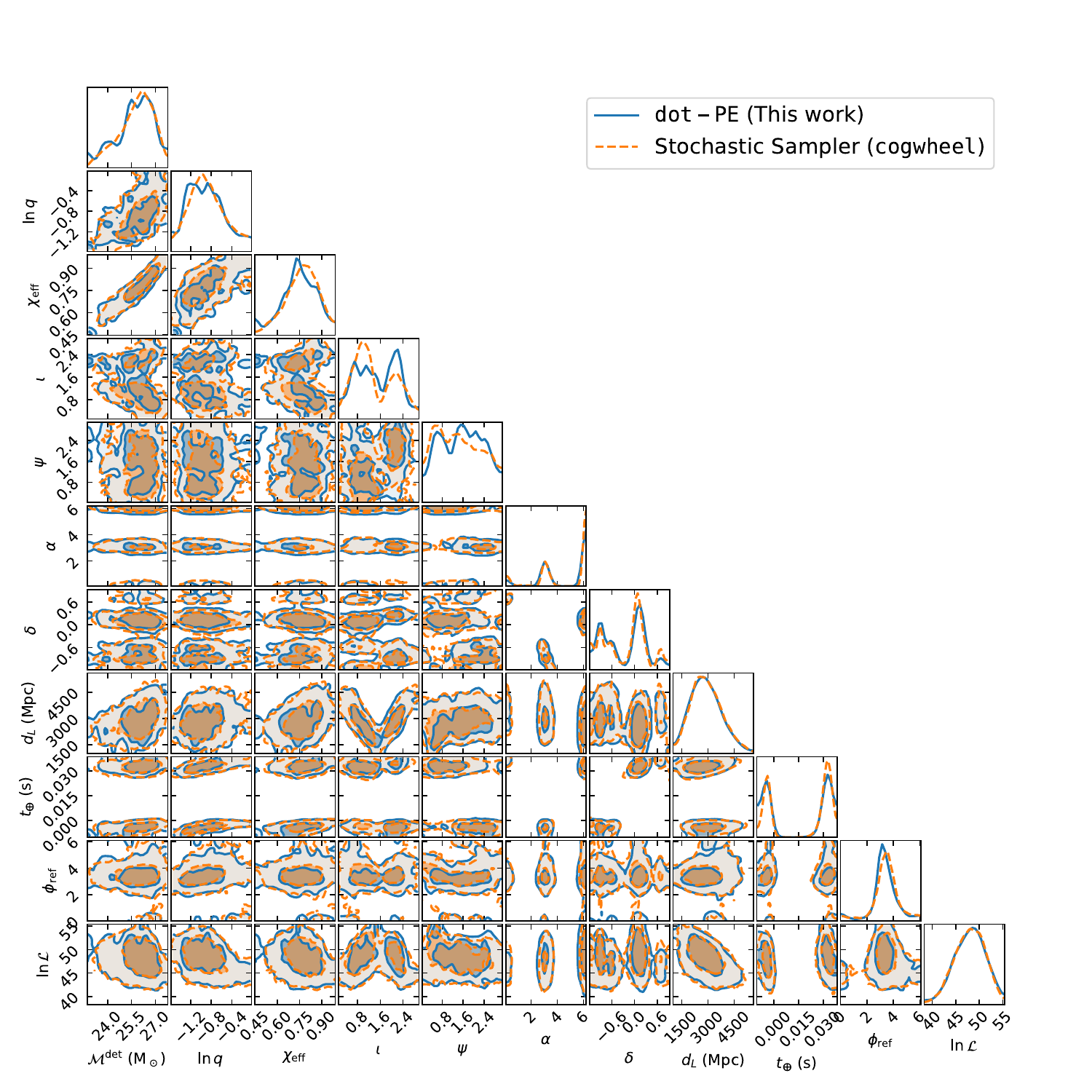}
    \caption[]{Comparison between the samples drawn from the posterior of Event 1 (see \ref{tab:single_events}) done with this work (blue) and with \texttt{cogwheel} using the extrinsic-marginalization method with \texttt{nautilus} (orange). The plot is produced with \texttt{cogwheel}'s plotting modules.}
    \label{fig:event_1}
\end{figure*}

\begin{figure*}
    \centering
    \includegraphics[width=\linewidth]{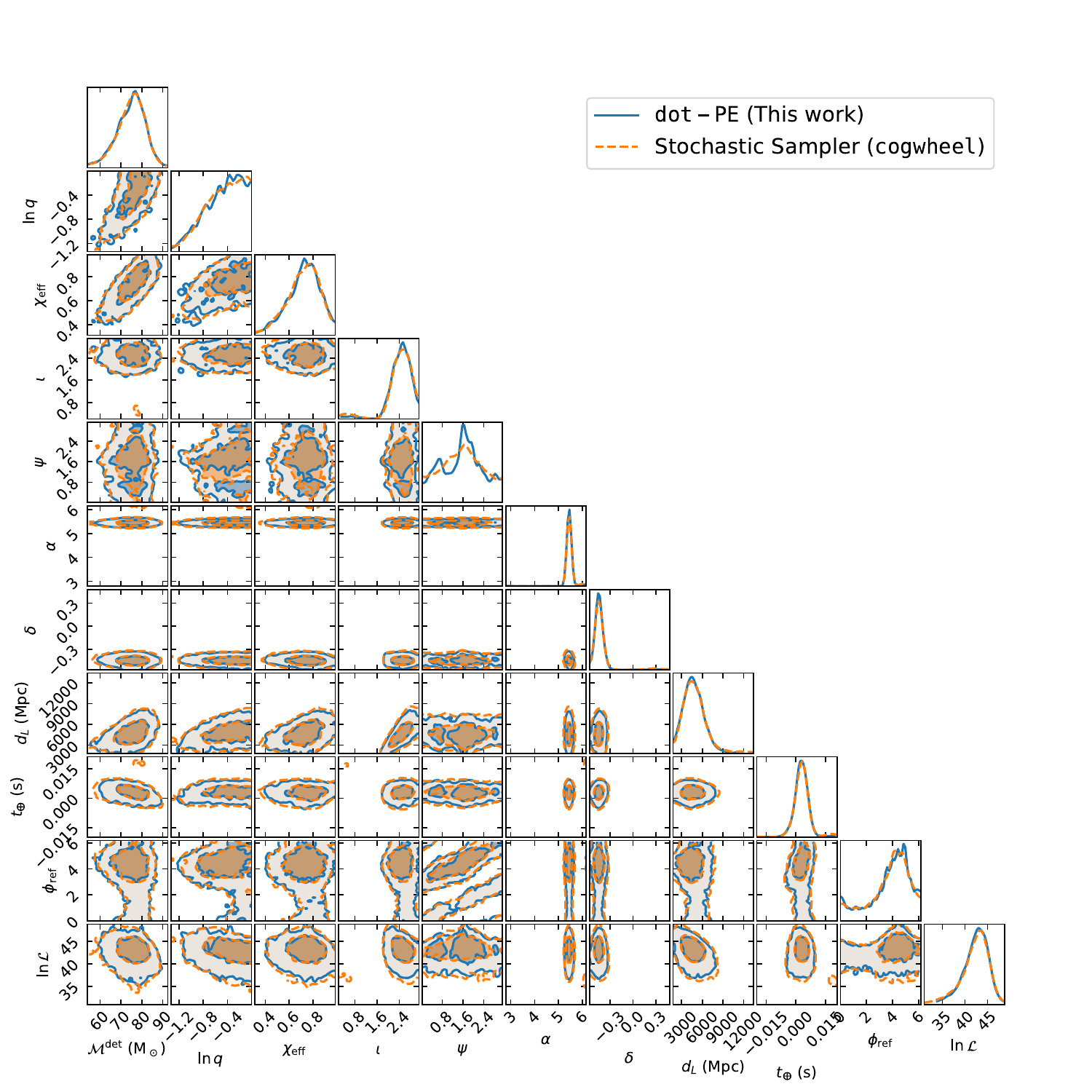}
    \caption[]{Same as Fig.~\ref{fig:event_1}, but with Event 2 (see Table \ref{tab:single_events}).}
    \label{fig:event_2}
\end{figure*}

\begin{figure*}
    \centering
    \includegraphics[width=\linewidth]{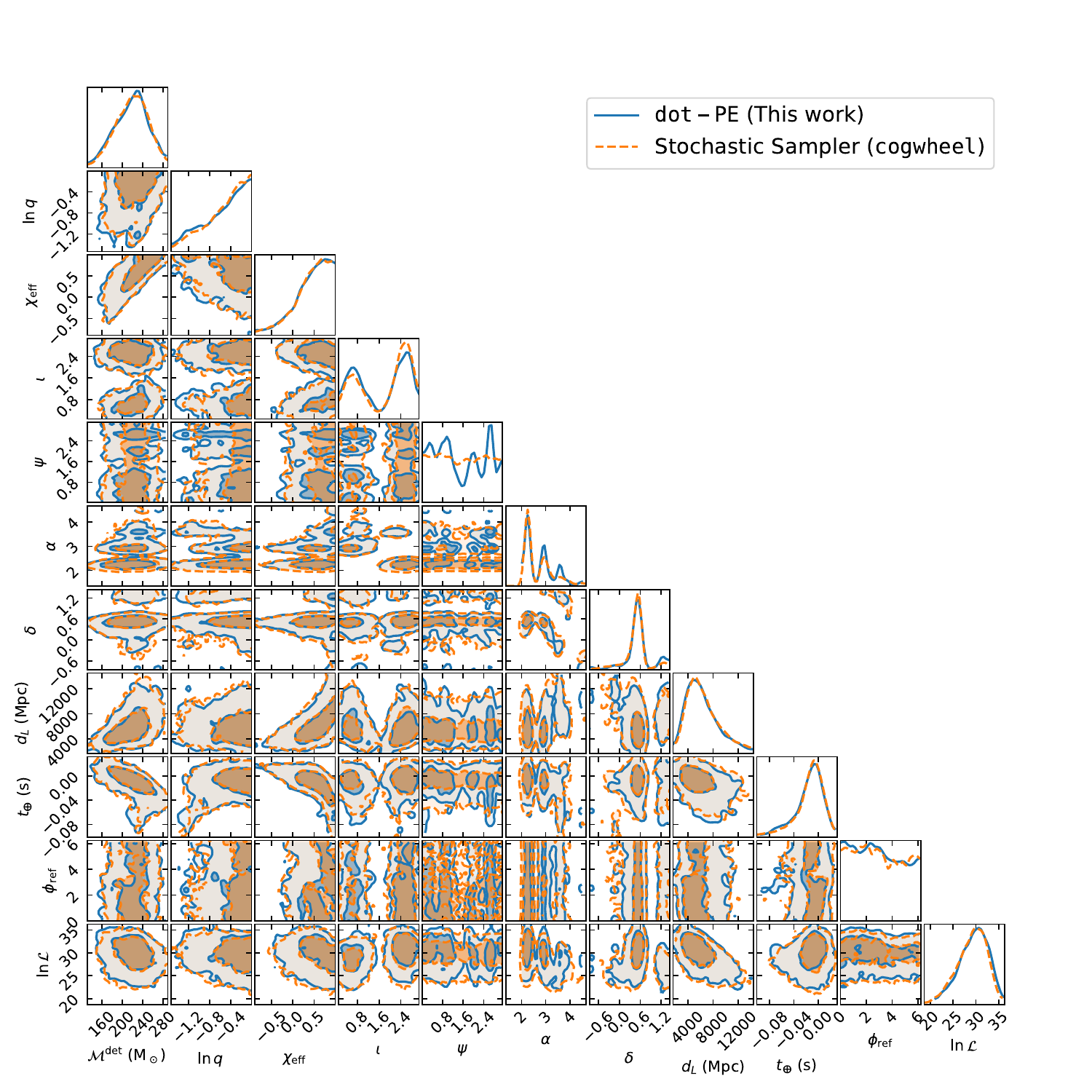}
    \caption[]{Same as Fig.~\ref{fig:event_1}, but with Event 3 (see Table \ref{tab:single_events}).}
    \label{fig:event_3}
\end{figure*}
\clearpage
\bibliographystyle{apsrev4-1-etal}
\bibliography{main}

\begin{thebibliography}{66}%
\makeatletter
\providecommand \@ifxundefined [1]{%
 \@ifx{#1\undefined}
}%
\providecommand \@ifnum [1]{%
 \ifnum #1\expandafter \@firstoftwo
 \else \expandafter \@secondoftwo
 \fi
}%
\providecommand \@ifx [1]{%
 \ifx #1\expandafter \@firstoftwo
 \else \expandafter \@secondoftwo
 \fi
}%
\providecommand \natexlab [1]{#1}%
\providecommand \enquote  [1]{``#1''}%
\providecommand \bibnamefont  [1]{#1}%
\providecommand \bibfnamefont [1]{#1}%
\providecommand \citenamefont [1]{#1}%
\providecommand \href@noop [0]{\@secondoftwo}%
\providecommand \href [0]{\begingroup \@sanitize@url \@href}%
\providecommand \@href[1]{\@@startlink{#1}\@@href}%
\providecommand \@@href[1]{\endgroup#1\@@endlink}%
\providecommand \@sanitize@url [0]{\catcode `\\12\catcode `\$12\catcode `\&12\catcode `\#12\catcode `\^12\catcode `\_12\catcode `\%12\relax}%
\providecommand \@@startlink[1]{}%
\providecommand \@@endlink[0]{}%
\providecommand \url  [0]{\begingroup\@sanitize@url \@url }%
\providecommand \@url [1]{\endgroup\@href {#1}{\urlprefix }}%
\providecommand \urlprefix  [0]{URL }%
\providecommand \Eprint [0]{\href }%
\providecommand \doibase [0]{http://dx.doi.org/}%
\providecommand \selectlanguage [0]{\@gobble}%
\providecommand \bibinfo  [0]{\@secondoftwo}%
\providecommand \bibfield  [0]{\@secondoftwo}%
\providecommand \translation [1]{[#1]}%
\providecommand \BibitemOpen [0]{}%
\providecommand \bibitemStop [0]{}%
\providecommand \bibitemNoStop [0]{.\EOS\space}%
\providecommand \EOS [0]{\spacefactor3000\relax}%
\providecommand \BibitemShut  [1]{\csname bibitem#1\endcsname}%
\let\auto@bib@innerbib\@empty
\bibitem [{\citenamefont {{LIGO Scientific Collaboration}}\ \emph {et~al.}(2015)\citenamefont {{LIGO Scientific Collaboration}}, \citenamefont {{Aasi}}, \citenamefont {{Abbott}}, \citenamefont {{Abbott}}, \citenamefont {{Abbott}}, \citenamefont {{Abernathy}}, \citenamefont {{Ackley}}, \citenamefont {{Adams}}, \citenamefont {{Adams}}, \citenamefont {{Addesso}}, \citenamefont {{Adhikari}}, \citenamefont {{Adya}}, \citenamefont {{Affeldt}}, \citenamefont {{Aggarwal}}, \citenamefont {{Aguiar}}, \citenamefont {{Ain}}, \citenamefont {{Ajith}}, \citenamefont {{Alemic}}, \citenamefont {{Allen}}, \citenamefont {{Amariutei}}, \citenamefont {{Anderson}}, \citenamefont {{Anderson}}, \citenamefont {{Arai}}, \citenamefont {{Araya}}, \citenamefont {{Arceneaux}}, \citenamefont {{Areeda}}, \citenamefont {{Ashton}}, \citenamefont {{Ast}}, \citenamefont {{Aston}}, \citenamefont {{Aufmuth}}, \citenamefont {{Aulbert}}, \citenamefont {{Aylott}}, \citenamefont {{Babak}}, \citenamefont {{Baker}}, \citenamefont {{Ballmer}},
  \citenamefont {{Barayoga}}, \citenamefont {{Barbet}}, \citenamefont {{Barclay}}, \citenamefont {{Barish}}, \citenamefont {{Barker}}, \citenamefont {{Barr}}, \citenamefont {{Barsotti}}, \citenamefont {{Bartlett}}, \citenamefont {{Barton}}, \citenamefont {{Bartos}}, \citenamefont {{Bassiri}}, \citenamefont {{Batch}}, \citenamefont {{Baune}}, \citenamefont {{Behnke}}, \citenamefont {{Bell}}, \citenamefont {{Bell}}, \citenamefont {{Benacquista}}, \citenamefont {{Bergman}}, \citenamefont {{Bergmann}}, \citenamefont {{Berry}}, \citenamefont {{Betzwieser}}, \citenamefont {{Bhagwat}}, \citenamefont {{Bhandare}}, \citenamefont {{Bilenko}}, \citenamefont {{Billingsley}}, \citenamefont {{Birch}}, \citenamefont {{Biscans}}, \citenamefont {{Biwer}}, \citenamefont {{Blackburn}}, \citenamefont {{Blackburn}}, \citenamefont {{Blair}}, \citenamefont {{Blair}}, \citenamefont {{Bock}}, \citenamefont {{Bodiya}}, \citenamefont {{Bojtos}}, \citenamefont {{Bond}}, \citenamefont {{Bork}}, \citenamefont {{Born}}, \citenamefont
  {{Bose}}, \citenamefont {{Brady}}, \citenamefont {{Braginsky}}, \citenamefont {{Brau}}, \citenamefont {{Bridges}}, \citenamefont {{Brinkmann}}, \citenamefont {{Brooks}}, \citenamefont {{Brown}}, \citenamefont {{Brown}}, \citenamefont {{Brown}}, \citenamefont {{Buchman}}, \citenamefont {{Buikema}}, \citenamefont {{Buonanno}}, \citenamefont {{Cadonati}}, \citenamefont {{Calder{\'o}n Bustillo}}, \citenamefont {{Camp}}, \citenamefont {{Cannon}}, \citenamefont {{Cao}}, \citenamefont {{Capano}}, \citenamefont {{Caride}}, \citenamefont {{Caudill}}, \citenamefont {{Cavagli{\`a}}}, \citenamefont {{Cepeda}}, \citenamefont {{Chakraborty}}, \citenamefont {{Chalermsongsak}}, \citenamefont {{Chamberlin}}, \citenamefont {{Chao}}, \citenamefont {{Charlton}}, \citenamefont {{Chen}}, \citenamefont {{Cho}}, \citenamefont {{Cho}}, \citenamefont {{Chow}}, \citenamefont {{Christensen}}, \citenamefont {{Chu}}, \citenamefont {{Chung}}, \citenamefont {{Ciani}}, \citenamefont {{Clara}}, \citenamefont {{Clark}}, \citenamefont
  {{Collette}}, \citenamefont {{Cominsky}}, \citenamefont {{Constancio}}, \citenamefont {{Cook}}, \citenamefont {{Corbitt}}, \citenamefont {{Cornish}}, \citenamefont {{Corsi}}, \citenamefont {{Costa}}, \citenamefont {{Coughlin}}, \citenamefont {{Countryman}}, \citenamefont {{Couvares}}, \citenamefont {{Coward}}, \citenamefont {{Cowart}}, \citenamefont {{Coyne}}, \citenamefont {{Coyne}}, \citenamefont {{Craig}}, \citenamefont {{Creighton}}, \citenamefont {{Creighton}}, \citenamefont {{Cripe}}, \citenamefont {{Crowder}}, \citenamefont {{Cumming}}, \citenamefont {{Cunningham}}, \citenamefont {{Cutler}}, \citenamefont {{Dahl}}, \citenamefont {{Dal Canton}}, \citenamefont {{Damjanic}}, \citenamefont {{Danilishin}}, \citenamefont {{Danzmann}}, \citenamefont {{Dartez}}, \citenamefont {{Dave}}, \citenamefont {{Daveloza}}, \citenamefont {{Davies}}, \citenamefont {{Daw}}, \citenamefont {{DeBra}}, \citenamefont {{Del Pozzo}}, \citenamefont {{Denker}}, \citenamefont {{Dent}}, \citenamefont {{Dergachev}}, \citenamefont
  {{DeRosa}}, \citenamefont {{DeSalvo}}, \citenamefont {{Dhurandhar}}, \citenamefont {{D{\textasciiacute}{\i}az}}, \citenamefont {{Di Palma}}, \citenamefont {{Dojcinoski}}, \citenamefont {{Dominguez}}, \citenamefont {{Donovan}}, \citenamefont {{Dooley}}, \citenamefont {{Doravari}}, \citenamefont {{Douglas}}, \citenamefont {{Downes}}, \citenamefont {{Driggers}}, \citenamefont {{Du}}, \citenamefont {{Dwyer}}, \citenamefont {{Eberle}}, \citenamefont {{Edo}}, \citenamefont {{Edwards}}, \citenamefont {{Edwards}}, \citenamefont {{Effler}}, \citenamefont {{Eggenstein}}, \citenamefont {{Ehrens}}, \citenamefont {{Eichholz}}, \citenamefont {{Eikenberry}}, \citenamefont {{Essick}}, \citenamefont {{Etzel}}, \citenamefont {{Evans}}, \citenamefont {{Evans}}, \citenamefont {{Factourovich}}, \citenamefont {{Fairhurst}}, \citenamefont {{Fan}}, \citenamefont {{Fang}}, \citenamefont {{Farr}}, \citenamefont {{Farr}}, \citenamefont {{Favata}}, \citenamefont {{Fays}}, \citenamefont {{Fehrmann}}, \citenamefont {{Fejer}},
  \citenamefont {{Feldbaum}}, \citenamefont {{Ferreira}}, \citenamefont {{Fisher}}, \citenamefont {{Frei}}, \citenamefont {{Freise}}, \citenamefont {{Frey}}, \citenamefont {{Fricke}}, \citenamefont {{Fritschel}}, \citenamefont {{Frolov}}, \citenamefont {{Fuentes-Tapia}}, \citenamefont {{Fulda}}, \citenamefont {{Fyffe}},\ and\ \citenamefont {{Gair}}}]{2015CQGra..32g4001L}%
  \BibitemOpen
  \bibfield  {author} {\bibinfo {author} {\bibnamefont {{LIGO Scientific Collaboration}}}, \bibinfo {author} {\bibfnamefont {J.}~\bibnamefont {{Aasi}}}, \bibinfo {author} {\bibfnamefont {B.~P.}\ \bibnamefont {{Abbott}}}, \bibinfo {author} {\bibfnamefont {R.}~\bibnamefont {{Abbott}}},  \emph {et~al.},\ }\href {\doibase 10.1088/0264-9381/32/7/074001} {\bibfield  {journal} {\bibinfo  {journal} {Classical and Quantum Gravity}\ }\textbf {\bibinfo {volume} {32}},\ \bibinfo {eid} {074001} (\bibinfo {year} {2015})},\ \Eprint {http://arxiv.org/abs/1411.4547} {arXiv:1411.4547 [gr-qc]} \BibitemShut {NoStop}%
\bibitem [{\citenamefont {Aasi}\ \emph {et~al.}(2012)\citenamefont {Aasi} \emph {et~al.}}]{Aasi2012}%
  \BibitemOpen
  \bibfield  {author} {\bibinfo {author} {\bibfnamefont {J.}~\bibnamefont {Aasi}} \emph {et~al.},\ }\href {\doibase 10.1088/0264-9381/29/15/155002} {\bibfield  {journal} {\bibinfo  {journal} {Classical and Quantum Gravity}\ }\textbf {\bibinfo {volume} {29}},\ \bibinfo {pages} {155002} (\bibinfo {year} {2012})}\BibitemShut {NoStop}%
\bibitem [{\citenamefont {{Abbott}}\ \emph {et~al.}(2023)\citenamefont {{Abbott}}, \citenamefont {{Abe}}, \citenamefont {{Acernese}}, \citenamefont {{Ackley}}, \citenamefont {{Adhicary}}, \citenamefont {{Adhikari}}, \citenamefont {{Adhikari}}, \citenamefont {{Adkins}}, \citenamefont {{Adya}}, \citenamefont {{Affeldt}}, \citenamefont {{Agarwal}}, \citenamefont {{Agathos}}, \citenamefont {{Aguiar}}, \citenamefont {{Aiello}}, \citenamefont {{Ain}}, \citenamefont {{Ajith}}, \citenamefont {{Akutsu}}, \citenamefont {{Albanesi}}, \citenamefont {{Alfaidi}}, \citenamefont {{Al-Jodah}}, \citenamefont {{All{\'e}n{\'e}}}, \citenamefont {{Allocca}}, \citenamefont {{Almualla}}, \citenamefont {{Altin}}, \citenamefont {{Amato}}, \citenamefont {{Amez-Droz}}, \citenamefont {{Amorosi}}, \citenamefont {{Anand}}, \citenamefont {{Ananyeva}}, \citenamefont {{Andersen}}, \citenamefont {{Anderson}}, \citenamefont {{Anderson}}, \citenamefont {{Andia}}, \citenamefont {{Ando}}, \citenamefont {{Andrade}}, \citenamefont {{Andres}},
  \citenamefont {{Andr{\'e}s-Carcasona}}, \citenamefont {{Andri{\'c}}}, \citenamefont {{Ansoldi}}, \citenamefont {{Antelis}}, \citenamefont {{Antier}}, \citenamefont {{Aoumi}}, \citenamefont {{Apostolatos}}, \citenamefont {{Appavuravther}}, \citenamefont {{Appert}}, \citenamefont {{Apple}}, \citenamefont {{Arai}}, \citenamefont {{Araya}}, \citenamefont {{Araya}}, \citenamefont {{Areeda}}, \citenamefont {{Ar{\`e}ne}}, \citenamefont {{Aritomi}}, \citenamefont {{Arnaud}}, \citenamefont {{Arogeti}}, \citenamefont {{Aronson}}, \citenamefont {{Arun}}, \citenamefont {{Asada}}, \citenamefont {{Ashton}}, \citenamefont {{Aso}}, \citenamefont {{Assiduo}}, \citenamefont {{Assis de Souza Melo}}, \citenamefont {{Aston}}, \citenamefont {{Astone}}, \citenamefont {{Aubin}}, \citenamefont {{Aultoneal}}, \citenamefont {{Babak}}, \citenamefont {{Badalyan}}, \citenamefont {{Badaracco}}, \citenamefont {{Badger}}, \citenamefont {{Bae}}, \citenamefont {{Bagnasco}}, \citenamefont {{Bai}}, \citenamefont {{Baier}}, \citenamefont
  {{Baiotti}}, \citenamefont {{Baird}}, \citenamefont {{Bajpai}}, \citenamefont {{Baka}}, \citenamefont {{Ball}}, \citenamefont {{Ballardin}}, \citenamefont {{Ballmer}}, \citenamefont {{Baltus}}, \citenamefont {{Banagiri}}, \citenamefont {{Banerjee}}, \citenamefont {{Bankar}}, \citenamefont {{Baral}}, \citenamefont {{Barayoga}}, \citenamefont {{Barber}}, \citenamefont {{Barish}}, \citenamefont {{Barker}}, \citenamefont {{Barneo}}, \citenamefont {{Barone}}, \citenamefont {{Barr}}, \citenamefont {{Barsotti}}, \citenamefont {{Barsuglia}}, \citenamefont {{Barta}}, \citenamefont {{Barthelmy}}, \citenamefont {{Barton}}, \citenamefont {{Bartos}}, \citenamefont {{Basak}}, \citenamefont {{Basalaev}}, \citenamefont {{Bassiri}}, \citenamefont {{Basti}}, \citenamefont {{Bawaj}}, \citenamefont {{Bayley}}, \citenamefont {{Baylor}}, \citenamefont {{Bazzan}}, \citenamefont {{B{\'e}csy}}, \citenamefont {{Bedakihale}}, \citenamefont {{Beirnaert}}, \citenamefont {{Bejger}}, \citenamefont {{Bell}}, \citenamefont {{Benedetto}},
  \citenamefont {{Beniwal}}, \citenamefont {{Benoit}}, \citenamefont {{Bentley}}, \citenamefont {{Yaala}}, \citenamefont {{Bera}}, \citenamefont {{Berbel}}, \citenamefont {{Bergamin}}, \citenamefont {{Berger}}, \citenamefont {{Bernuzzi}}, \citenamefont {{Beroiz}}, \citenamefont {{Berry}}, \citenamefont {{Bersanetti}}, \citenamefont {{Bertolini}}, \citenamefont {{Betzwieser}}, \citenamefont {{Beveridge}}, \citenamefont {{Bevins}}, \citenamefont {{Bhandare}}, \citenamefont {{Bhandari}}, \citenamefont {{Bhardwaj}}, \citenamefont {{Bhatt}}, \citenamefont {{Bhattacharjee}}, \citenamefont {{Bhaumik}}, \citenamefont {{Bianchi}}, \citenamefont {{Bilenko}}, \citenamefont {{Bilicki}}, \citenamefont {{Billingsley}}, \citenamefont {{Bini}}, \citenamefont {{Birnholtz}}, \citenamefont {{Biscans}}, \citenamefont {{Bischi}}, \citenamefont {{Biscoveanu}}, \citenamefont {{Bisht}}, \citenamefont {{Biswas}}, \citenamefont {{Bitossi}}, \citenamefont {{Bizouard}}, \citenamefont {{Blackburn}}, \citenamefont {{Blair}}, \citenamefont
  {{Blair}}, \citenamefont {{Blair}}, \citenamefont {{Bobba}}, \citenamefont {{Bode}}, \citenamefont {{Bo{\"e}r}}, \citenamefont {{Bogaert}}, \citenamefont {{Boileau}}, \citenamefont {{Boldrini}}, \citenamefont {{Bolingbroke}}, \citenamefont {{Bonavena}}, \citenamefont {{Bondarescu}}, \citenamefont {{Bondu}}, \citenamefont {{Bonilla}}, \citenamefont {{Bonilla}}, \citenamefont {{Bonnand}}, \citenamefont {{Booker}}, \citenamefont {{Bork}}, \citenamefont {{Boschi}}, \citenamefont {{Bose}}, \citenamefont {{Bose}}, \citenamefont {{Bossilkov}}, \citenamefont {{Boudart}}, \citenamefont {{Bouffanais}}, \citenamefont {{Bozzi}}, \citenamefont {{Bradaschia}}, \citenamefont {{Brady}}, \citenamefont {{Braglia}}, \citenamefont {{Branch}}, \citenamefont {{Branchesi}}, \citenamefont {{Brau}}, \citenamefont {{Breschi}}, \citenamefont {{Briant}}, \citenamefont {{Brillet}}, \citenamefont {{Brinkmann}}, \citenamefont {{Brockill}}, \citenamefont {{Brooks}}, \citenamefont {{Brooks}}, \citenamefont {{Brown}}, \citenamefont
  {{Brunett}}, \citenamefont {{Bruno}}, \citenamefont {{Bruntz}}, \citenamefont {{Bryant}}, \citenamefont {{Bucci}}, \citenamefont {{Buchanan}}, \citenamefont {{Bulashenko}}, \citenamefont {{Bulik}}, \citenamefont {{Bulten}}, \citenamefont {{Buonanno}}, \citenamefont {{Burtnyk}}, \citenamefont {{Buscicchio}},\ and\ \citenamefont {{Buskulic}}}]{GWOSC2023}%
  \BibitemOpen
  \bibfield  {author} {\bibinfo {author} {\bibfnamefont {R.}~\bibnamefont {{Abbott}}}, \bibinfo {author} {\bibfnamefont {H.}~\bibnamefont {{Abe}}}, \bibinfo {author} {\bibfnamefont {F.}~\bibnamefont {{Acernese}}}, \bibinfo {author} {\bibfnamefont {K.}~\bibnamefont {{Ackley}}},  \emph {et~al.},\ }\href {\doibase 10.3847/1538-4365/acdc9f} {\bibfield  {journal} {\bibinfo  {journal} {The Astrophysical Journal Supplement Series}\ }\textbf {\bibinfo {volume} {267}},\ \bibinfo {eid} {29} (\bibinfo {year} {2023})},\ \Eprint {http://arxiv.org/abs/2302.03676} {arXiv:2302.03676 [gr-qc]} \BibitemShut {NoStop}%
\bibitem [{\citenamefont {Wadekar}\ \emph {et~al.}(2023)\citenamefont {Wadekar}, \citenamefont {Roulet}, \citenamefont {Venumadhav}, \citenamefont {Mehta}, \citenamefont {Zackay}, \citenamefont {Mushkin}, \citenamefont {Olsen},\ and\ \citenamefont {Zaldarriaga}}]{Wadekar2023}%
  \BibitemOpen
  \bibfield  {author} {\bibinfo {author} {\bibfnamefont {D.}~\bibnamefont {Wadekar}}, \bibinfo {author} {\bibfnamefont {J.}~\bibnamefont {Roulet}}, \bibinfo {author} {\bibfnamefont {T.}~\bibnamefont {Venumadhav}}, \bibinfo {author} {\bibfnamefont {A.~K.}\ \bibnamefont {Mehta}},  \emph {et~al.},\ }\href@noop {} {\enquote {\bibinfo {title} {New black hole mergers in the {LIGO}--{V}irgo {O3} data from a gravitational wave search including higher-order harmonics},}\ } (\bibinfo {year} {2023}),\ \Eprint {http://arxiv.org/abs/2312.06631} {arXiv:2312.06631 [gr-qc]} \BibitemShut {NoStop}%
\bibitem [{\citenamefont {Abbott}\ \emph {et~al.}(2023)\citenamefont {Abbott} \emph {et~al.}}]{Abbott2023_GWTC3}%
  \BibitemOpen
  \bibfield  {author} {\bibinfo {author} {\bibfnamefont {R.}~\bibnamefont {Abbott}} \emph {et~al.} (\bibinfo {collaboration} {LIGO Scientific Collaboration, Virgo Collaboration, and KAGRA Collaboration}),\ }\href {\doibase 10.1103/PhysRevX.13.041039} {\bibfield  {journal} {\bibinfo  {journal} {Phys. Rev. X}\ }\textbf {\bibinfo {volume} {13}},\ \bibinfo {pages} {041039} (\bibinfo {year} {2023})}\BibitemShut {NoStop}%
\bibitem [{\citenamefont {{Nitz}}\ \emph {et~al.}(2023)\citenamefont {{Nitz}}, \citenamefont {{Kumar}}, \citenamefont {{Wang}}, \citenamefont {{Kastha}}, \citenamefont {{Wu}}, \citenamefont {{Sch{\"a}fer}}, \citenamefont {{Dhurkunde}},\ and\ \citenamefont {{Capano}}}]{Nitz4OGC2023}%
  \BibitemOpen
  \bibfield  {author} {\bibinfo {author} {\bibfnamefont {A.~H.}\ \bibnamefont {{Nitz}}}, \bibinfo {author} {\bibfnamefont {S.}~\bibnamefont {{Kumar}}}, \bibinfo {author} {\bibfnamefont {Y.-F.}\ \bibnamefont {{Wang}}}, \bibinfo {author} {\bibfnamefont {S.}~\bibnamefont {{Kastha}}},  \emph {et~al.},\ }\href {\doibase 10.3847/1538-4357/aca591} {\bibfield  {journal} {\bibinfo  {journal} {\apj}\ }\textbf {\bibinfo {volume} {946}},\ \bibinfo {eid} {59} (\bibinfo {year} {2023})},\ \Eprint {http://arxiv.org/abs/2112.06878} {arXiv:2112.06878 [astro-ph.HE]} \BibitemShut {NoStop}%
\bibitem [{\citenamefont {{Capote}}\ \emph {et~al.}(2025)\citenamefont {{Capote}}, \citenamefont {{Jia}}, \citenamefont {{Aritomi}}, \citenamefont {{Nakano}}, \citenamefont {{Xu}}, \citenamefont {{Abbott}}, \citenamefont {{Abouelfettouh}}, \citenamefont {{Adhikari}}, \citenamefont {{Ananyeva}}, \citenamefont {{Appert}}, \citenamefont {{Apple}}, \citenamefont {{Arai}}, \citenamefont {{Aston}}, \citenamefont {{Ball}}, \citenamefont {{Ballmer}}, \citenamefont {{Barker}}, \citenamefont {{Barsotti}}, \citenamefont {{Berger}}, \citenamefont {{Betzwieser}}, \citenamefont {{Bhattacharjee}}, \citenamefont {{Billingsley}}, \citenamefont {{Biscans}}, \citenamefont {{Blair}}, \citenamefont {{Bode}}, \citenamefont {{Bonilla}}, \citenamefont {{Bossilkov}}, \citenamefont {{Branch}}, \citenamefont {{Brooks}}, \citenamefont {{Brown}}, \citenamefont {{Bryant}}, \citenamefont {{Cahillane}}, \citenamefont {{Cao}}, \citenamefont {{Clara}}, \citenamefont {{Collins}}, \citenamefont {{Compton}}, \citenamefont {{Cottingham}},
  \citenamefont {{Coyne}}, \citenamefont {{Crouch}}, \citenamefont {{Csizmazia}}, \citenamefont {{Cumming}}, \citenamefont {{Dartez}}, \citenamefont {{Davis}}, \citenamefont {{Demos}}, \citenamefont {{Dohmen}}, \citenamefont {{Driggers}}, \citenamefont {{Dwyer}}, \citenamefont {{Effler}}, \citenamefont {{Ejlli}}, \citenamefont {{Etzel}}, \citenamefont {{Evans}}, \citenamefont {{Feicht}}, \citenamefont {{Frey}}, \citenamefont {{Frischhertz}}, \citenamefont {{Fritschel}}, \citenamefont {{Frolov}}, \citenamefont {{Fuentes-Garcia}}, \citenamefont {{Fulda}}, \citenamefont {{Fyffe}}, \citenamefont {{Ganapathy}}, \citenamefont {{Gateley}}, \citenamefont {{Gayer}}, \citenamefont {{Giaime}}, \citenamefont {{Giardina}}, \citenamefont {{Glanzer}}, \citenamefont {{Goetz}}, \citenamefont {{Goetz}}, \citenamefont {{Goodwin-Jones}}, \citenamefont {{Gras}}, \citenamefont {{Gray}}, \citenamefont {{Griffith}}, \citenamefont {{Grote}}, \citenamefont {{Guidry}}, \citenamefont {{Gurs}}, \citenamefont {{Hall}}, \citenamefont
  {{Hanks}}, \citenamefont {{Hanson}}, \citenamefont {{Heintze}}, \citenamefont {{Helmling-Cornell}}, \citenamefont {{Holland}}, \citenamefont {{Hoyland}}, \citenamefont {{Huang}}, \citenamefont {{Inoue}}, \citenamefont {{James}}, \citenamefont {{Jamies}}, \citenamefont {{Jennings}}, \citenamefont {{Jones}}, \citenamefont {{Kabagoz}}, \citenamefont {{Karat}}, \citenamefont {{Karki}}, \citenamefont {{Kasprzack}}, \citenamefont {{Kawabe}}, \citenamefont {{Kijbunchoo}}, \citenamefont {{King}}, \citenamefont {{Kissel}}, \citenamefont {{Komori}}, \citenamefont {{Kontos}}, \citenamefont {{Kumar}}, \citenamefont {{Kuns}}, \citenamefont {{Landry}}, \citenamefont {{Lantz}}, \citenamefont {{Laxen}}, \citenamefont {{Lee}}, \citenamefont {{Lesovsky}}, \citenamefont {{Villarreal}}, \citenamefont {{Lormand}}, \citenamefont {{Loughlin}}, \citenamefont {{Macas}}, \citenamefont {{MacInnis}}, \citenamefont {{Makarem}}, \citenamefont {{Mannix}}, \citenamefont {{Mansell}}, \citenamefont {{Martin}}, \citenamefont {{Mason}},
  \citenamefont {{Matichard}}, \citenamefont {{Mavalvala}}, \citenamefont {{Maxwell}}, \citenamefont {{McCarrol}}, \citenamefont {{McCarthy}}, \citenamefont {{McClelland}}, \citenamefont {{McCormick}}, \citenamefont {{McRae}}, \citenamefont {{Mera}}, \citenamefont {{Merilh}}, \citenamefont {{Meylahn}}, \citenamefont {{Mittleman}}, \citenamefont {{Moraru}}, \citenamefont {{Moreno}}, \citenamefont {{Mullavey}}, \citenamefont {{Nelson}}, \citenamefont {{Neunzert}}, \citenamefont {{Notte}}, \citenamefont {{Oberling}}, \citenamefont {{O'Hanlon}}, \citenamefont {{Osthelder}}, \citenamefont {{Ottaway}}, \citenamefont {{Overmier}}, \citenamefont {{Parker}}, \citenamefont {{Patane}}, \citenamefont {{Pele}}, \citenamefont {{Pham}}, \citenamefont {{Pirello}}, \citenamefont {{Pullin}}, \citenamefont {{Quetschke}}, \citenamefont {{Ramirez}}, \citenamefont {{Ransom}}, \citenamefont {{Reyes}}, \citenamefont {{Richardson}}, \citenamefont {{Robinson}}, \citenamefont {{Rollins}}, \citenamefont {{Romel}}, \citenamefont
  {{Romie}}, \citenamefont {{Ross}}, \citenamefont {{Ryan}}, \citenamefont {{Sadecki}}, \citenamefont {{Sanchez}}, \citenamefont {{Sanchez}}, \citenamefont {{Sanchez}}, \citenamefont {{Savage}}, \citenamefont {{Schaetzl}}, \citenamefont {{Schiworski}}, \citenamefont {{Schnabel}}, \citenamefont {{Schofield}}, \citenamefont {{Schwartz}}, \citenamefont {{Sellers}}, \citenamefont {{Shaffer}}, \citenamefont {{Short}}, \citenamefont {{Sigg}}, \citenamefont {{Slagmolen}}, \citenamefont {{Soike}}, \citenamefont {{Soni}}, \citenamefont {{Srivastava}}, \citenamefont {{Sun}}, \citenamefont {{Tanner}}, \citenamefont {{Thomas}}, \citenamefont {{Thomas}}, \citenamefont {{Thorne}}, \citenamefont {{Todd}}, \citenamefont {{Torrie}}, \citenamefont {{Traylor}}, \citenamefont {{Ubhi}}, \citenamefont {{Vajente}}, \citenamefont {{Vanosky}}, \citenamefont {{Vecchio}}, \citenamefont {{Veitch}}, \citenamefont {{Vibhute}}, \citenamefont {{von Reis}}, \citenamefont {{Warner}}, \citenamefont {{Weaver}}, \citenamefont {{Weiss}},
  \citenamefont {{Whittle}}, \citenamefont {{Willke}}, \citenamefont {{Wipf}}, \citenamefont {{Wright}}, \citenamefont {{Yamamoto}}, \citenamefont {{Zhang}},\ and\ \citenamefont {{Zucker}}}]{CapoteO4performance2025}%
  \BibitemOpen
  \bibfield  {author} {\bibinfo {author} {\bibfnamefont {E.}~\bibnamefont {{Capote}}}, \bibinfo {author} {\bibfnamefont {W.}~\bibnamefont {{Jia}}}, \bibinfo {author} {\bibfnamefont {N.}~\bibnamefont {{Aritomi}}}, \bibinfo {author} {\bibfnamefont {M.}~\bibnamefont {{Nakano}}},  \emph {et~al.},\ }\href {\doibase 10.1103/PhysRevD.111.062002} {\bibfield  {journal} {\bibinfo  {journal} {\prd}\ }\textbf {\bibinfo {volume} {111}},\ \bibinfo {eid} {062002} (\bibinfo {year} {2025})},\ \Eprint {http://arxiv.org/abs/2411.14607} {arXiv:2411.14607 [gr-qc]} \BibitemShut {NoStop}%
\bibitem [{\citenamefont {{LIGO Laboratory}}(2025)}]{LIGO2025O4Milestone}%
  \BibitemOpen
  \bibfield  {author} {\bibinfo {author} {\bibnamefont {{LIGO Laboratory}}},\ }\href@noop {} {\enquote {\bibinfo {title} {{LIGO} detects 200th gravitational-wave candidate in {O}4},}\ }\bibinfo {howpublished} {\url{https://www.ligo.caltech.edu/MIT/news/ligo20250320}} (\bibinfo {year} {2025}),\ \bibinfo {note} {accessed: 2025-06-05}\BibitemShut {NoStop}%
\bibitem [{\citenamefont {{Abac}}\ \emph {et~al.}(2025)\citenamefont {{Abac}}, \citenamefont {{Abramo}}, \citenamefont {{Albanesi}}, \citenamefont {{Albertini}}, \citenamefont {{Agapito}}, \citenamefont {{Agathos}} \emph {et~al.}}]{Abac2025ET}%
  \BibitemOpen
  \bibfield  {author} {\bibinfo {author} {\bibfnamefont {A.}~\bibnamefont {{Abac}}}, \bibinfo {author} {\bibfnamefont {R.}~\bibnamefont {{Abramo}}}, \bibinfo {author} {\bibfnamefont {S.}~\bibnamefont {{Albanesi}}}, \bibinfo {author} {\bibfnamefont {A.}~\bibnamefont {{Albertini}}}, \bibinfo {author} {\bibfnamefont {A.}~\bibnamefont {{Agapito}}}, \bibinfo {author} {\bibfnamefont {M.}~\bibnamefont {{Agathos}}},  \emph {et~al.},\ }\href {\doibase 10.48550/arXiv.2503.12263} {\bibfield  {journal} {\bibinfo  {journal} {arXiv e-prints}\ ,\ \bibinfo {eid} {arXiv:2503.12263}} (\bibinfo {year} {2025})},\ \Eprint {http://arxiv.org/abs/2503.12263} {arXiv:2503.12263 [gr-qc]} \BibitemShut {NoStop}%
\bibitem [{\citenamefont {{Evans}}\ \emph {et~al.}(2021)\citenamefont {{Evans}}, \citenamefont {{Adhikari}}, \citenamefont {{Afle}}, \citenamefont {{Ballmer}}, \citenamefont {{Biscoveanu}}, \citenamefont {{Borhanian}}, \citenamefont {{Brown}}, \citenamefont {{Chen}}, \citenamefont {{Eisenstein}}, \citenamefont {{Gruson}}, \citenamefont {{Gupta}}, \citenamefont {{Hall}}, \citenamefont {{Huxford}}, \citenamefont {{Kamai}}, \citenamefont {{Kashyap}}, \citenamefont {{Kissel}}, \citenamefont {{Kuns}}, \citenamefont {{Landry}}, \citenamefont {{Lenon}}, \citenamefont {{Lovelace}}, \citenamefont {{McCuller}}, \citenamefont {{Ng}}, \citenamefont {{Nitz}}, \citenamefont {{Read}}, \citenamefont {{Sathyaprakash}}, \citenamefont {{Shoemaker}}, \citenamefont {{Slagmolen}}, \citenamefont {{Smith}}, \citenamefont {{Srivastava}}, \citenamefont {{Sun}}, \citenamefont {{Vitale}},\ and\ \citenamefont {{Weiss}}}]{Evans2021}%
  \BibitemOpen
  \bibfield  {author} {\bibinfo {author} {\bibfnamefont {M.}~\bibnamefont {{Evans}}}, \bibinfo {author} {\bibfnamefont {R.~X.}\ \bibnamefont {{Adhikari}}}, \bibinfo {author} {\bibfnamefont {C.}~\bibnamefont {{Afle}}}, \bibinfo {author} {\bibfnamefont {S.~W.}\ \bibnamefont {{Ballmer}}},  \emph {et~al.},\ }\href {\doibase 10.48550/arXiv.2109.09882} {\bibfield  {journal} {\bibinfo  {journal} {arXiv e-prints}\ ,\ \bibinfo {eid} {arXiv:2109.09882}} (\bibinfo {year} {2021})},\ \Eprint {http://arxiv.org/abs/2109.09882} {arXiv:2109.09882 [astro-ph.IM]} \BibitemShut {NoStop}%
\bibitem [{\citenamefont {{LIGO Scientific Collaboration}}(2024)}]{LIGO-T2200287}%
  \BibitemOpen
  \bibfield  {author} {\bibinfo {author} {\bibnamefont {{LIGO Scientific Collaboration}}},\ }\href {https://dcc.ligo.org/public/0183/T2200287/003/T2200287v3_PO5report.pdf} {\emph {\bibinfo {title} {{Report from the LSC Post-O5 Study Group}}}},\ \bibinfo {type} {Tech. Rep.}\ \bibinfo {number} {LIGO-T2200287–v3}\ (\bibinfo  {institution} {{LIGO Document Control Center}},\ \bibinfo {year} {2024})\BibitemShut {NoStop}%
\bibitem [{\citenamefont {{Saleem}}\ \emph {et~al.}(2022)\citenamefont {{Saleem}}, \citenamefont {{Rana}}, \citenamefont {{Gayathri}}, \citenamefont {{Vijaykumar}}, \citenamefont {{Goyal}}, \citenamefont {{Sachdev}}, \citenamefont {{Suresh}}, \citenamefont {{Sudhagar}}, \citenamefont {{Mukherjee}}, \citenamefont {{Gaur}}, \citenamefont {{Sathyaprakash}}, \citenamefont {{Pai}}, \citenamefont {{Adhikari}}, \citenamefont {{Ajith}},\ and\ \citenamefont {{Bose}}}]{Saleem2022LIGOindia}%
  \BibitemOpen
  \bibfield  {author} {\bibinfo {author} {\bibfnamefont {M.}~\bibnamefont {{Saleem}}}, \bibinfo {author} {\bibfnamefont {J.}~\bibnamefont {{Rana}}}, \bibinfo {author} {\bibfnamefont {V.}~\bibnamefont {{Gayathri}}}, \bibinfo {author} {\bibfnamefont {A.}~\bibnamefont {{Vijaykumar}}},  \emph {et~al.},\ }\href {\doibase 10.1088/1361-6382/ac3b99} {\bibfield  {journal} {\bibinfo  {journal} {Classical and Quantum Gravity}\ }\textbf {\bibinfo {volume} {39}},\ \bibinfo {eid} {025004} (\bibinfo {year} {2022})},\ \Eprint {http://arxiv.org/abs/2105.01716} {arXiv:2105.01716 [gr-qc]} \BibitemShut {NoStop}%
\bibitem [{\citenamefont {Farr}(2014)}]{Farr2014}%
  \BibitemOpen
  \bibfield  {author} {\bibinfo {author} {\bibfnamefont {W.}~\bibnamefont {Farr}},\ }\href@noop {} {\enquote {\bibinfo {title} {Marginalisation of the time and phase parameters in {CBC} parameter estimation},}\ }\bibinfo {howpublished} {\url{https://dcc.ligo.org/public/0114/T1400460/002/margtime.pdf}} (\bibinfo {year} {2014})\BibitemShut {NoStop}%
\bibitem [{\citenamefont {{Farr}}\ \emph {et~al.}(2014)\citenamefont {{Farr}}, \citenamefont {{Ochsner}}, \citenamefont {{Farr}},\ and\ \citenamefont {{O'Shaughnessy}}}]{Farr2014PhRvD}%
  \BibitemOpen
  \bibfield  {author} {\bibinfo {author} {\bibfnamefont {B.}~\bibnamefont {{Farr}}}, \bibinfo {author} {\bibfnamefont {E.}~\bibnamefont {{Ochsner}}}, \bibinfo {author} {\bibfnamefont {W.~M.}\ \bibnamefont {{Farr}}}, \ and\ \bibinfo {author} {\bibfnamefont {R.}~\bibnamefont {{O'Shaughnessy}}},\ }\href {\doibase 10.1103/PhysRevD.90.024018} {\bibfield  {journal} {\bibinfo  {journal} {\prd}\ }\textbf {\bibinfo {volume} {90}},\ \bibinfo {eid} {024018} (\bibinfo {year} {2014})},\ \Eprint {http://arxiv.org/abs/1404.7070} {arXiv:1404.7070 [gr-qc]} \BibitemShut {NoStop}%
\bibitem [{\citenamefont {Veitch}\ \emph {et~al.}(2015)\citenamefont {Veitch}, \citenamefont {Raymond}, \citenamefont {Farr}, \citenamefont {Farr}, \citenamefont {Graff}, \citenamefont {Vitale}, \citenamefont {Aylott}, \citenamefont {Blackburn}, \citenamefont {Christensen}, \citenamefont {Coughlin}, \citenamefont {Del~Pozzo}, \citenamefont {Feroz}, \citenamefont {Gair}, \citenamefont {Haster}, \citenamefont {Kalogera}, \citenamefont {Littenberg}, \citenamefont {Mandel}, \citenamefont {O’Shaughnessy}, \citenamefont {Pitkin}, \citenamefont {Rodriguez}, \citenamefont {R\"{o}ver}, \citenamefont {Sidery}, \citenamefont {Smith}, \citenamefont {Van Der~Sluys}, \citenamefont {Vecchio}, \citenamefont {Vousden},\ and\ \citenamefont {Wade}}]{Veitch2015}%
  \BibitemOpen
  \bibfield  {author} {\bibinfo {author} {\bibfnamefont {J.}~\bibnamefont {Veitch}}, \bibinfo {author} {\bibfnamefont {V.}~\bibnamefont {Raymond}}, \bibinfo {author} {\bibfnamefont {B.}~\bibnamefont {Farr}}, \bibinfo {author} {\bibfnamefont {W.}~\bibnamefont {Farr}},  \emph {et~al.},\ }\href {\doibase 10.1103/physrevd.91.042003} {\bibfield  {journal} {\bibinfo  {journal} {Physical Review D}\ }\textbf {\bibinfo {volume} {91}} (\bibinfo {year} {2015}),\ 10.1103/physrevd.91.042003}\BibitemShut {NoStop}%
\bibitem [{\citenamefont {Pankow}\ \emph {et~al.}(2015)\citenamefont {Pankow}, \citenamefont {Brady}, \citenamefont {Ochsner},\ and\ \citenamefont {O’Shaughnessy}}]{Pankow2015}%
  \BibitemOpen
  \bibfield  {author} {\bibinfo {author} {\bibfnamefont {C.}~\bibnamefont {Pankow}}, \bibinfo {author} {\bibfnamefont {P.}~\bibnamefont {Brady}}, \bibinfo {author} {\bibfnamefont {E.}~\bibnamefont {Ochsner}}, \ and\ \bibinfo {author} {\bibfnamefont {R.}~\bibnamefont {O’Shaughnessy}},\ }\href {\doibase 10.1103/physrevd.92.023002} {\bibfield  {journal} {\bibinfo  {journal} {Physical Review D}\ }\textbf {\bibinfo {volume} {92}} (\bibinfo {year} {2015}),\ 10.1103/physrevd.92.023002}\BibitemShut {NoStop}%
\bibitem [{\citenamefont {Zackay}\ \emph {et~al.}(2018)\citenamefont {Zackay}, \citenamefont {Dai},\ and\ \citenamefont {Venumadhav}}]{Zackay2018}%
  \BibitemOpen
  \bibfield  {author} {\bibinfo {author} {\bibfnamefont {B.}~\bibnamefont {Zackay}}, \bibinfo {author} {\bibfnamefont {L.}~\bibnamefont {Dai}}, \ and\ \bibinfo {author} {\bibfnamefont {T.}~\bibnamefont {Venumadhav}},\ }\href@noop {} {\enquote {\bibinfo {title} {Relative binning and fast likelihood evaluation for gravitational wave parameter estimation},}\ } (\bibinfo {year} {2018}),\ \Eprint {http://arxiv.org/abs/1806.08792} {arXiv:1806.08792 [astro-ph.IM]} \BibitemShut {NoStop}%
\bibitem [{\citenamefont {Biwer}\ \emph {et~al.}(2019)\citenamefont {Biwer}, \citenamefont {Capano}, \citenamefont {De}, \citenamefont {Cabero}, \citenamefont {Brown}, \citenamefont {Nitz},\ and\ \citenamefont {Raymond}}]{Biwer2019}%
  \BibitemOpen
  \bibfield  {author} {\bibinfo {author} {\bibfnamefont {C.~M.}\ \bibnamefont {Biwer}}, \bibinfo {author} {\bibfnamefont {C.~D.}\ \bibnamefont {Capano}}, \bibinfo {author} {\bibfnamefont {S.}~\bibnamefont {De}}, \bibinfo {author} {\bibfnamefont {M.}~\bibnamefont {Cabero}}, \bibinfo {author} {\bibfnamefont {D.~A.}\ \bibnamefont {Brown}}, \bibinfo {author} {\bibfnamefont {A.~H.}\ \bibnamefont {Nitz}}, \ and\ \bibinfo {author} {\bibfnamefont {V.}~\bibnamefont {Raymond}},\ }\href {\doibase 10.1088/1538-3873/aaef0b} {\bibfield  {journal} {\bibinfo  {journal} {Publications of the Astronomical Society of the Pacific}\ }\textbf {\bibinfo {volume} {131}},\ \bibinfo {pages} {024503} (\bibinfo {year} {2019})}\BibitemShut {NoStop}%
\bibitem [{\citenamefont {Dax}\ \emph {et~al.}(2021)\citenamefont {Dax}, \citenamefont {Green}, \citenamefont {Gair}, \citenamefont {Macke}, \citenamefont {Buonanno},\ and\ \citenamefont {Sch\"olkopf}}]{Dax2021}%
  \BibitemOpen
  \bibfield  {author} {\bibinfo {author} {\bibfnamefont {M.}~\bibnamefont {Dax}}, \bibinfo {author} {\bibfnamefont {S.~R.}\ \bibnamefont {Green}}, \bibinfo {author} {\bibfnamefont {J.}~\bibnamefont {Gair}}, \bibinfo {author} {\bibfnamefont {J.~H.}\ \bibnamefont {Macke}}, \bibinfo {author} {\bibfnamefont {A.}~\bibnamefont {Buonanno}}, \ and\ \bibinfo {author} {\bibfnamefont {B.}~\bibnamefont {Sch\"olkopf}},\ }\href {\doibase 10.1103/PhysRevLett.127.241103} {\bibfield  {journal} {\bibinfo  {journal} {Phys. Rev. Lett.}\ }\textbf {\bibinfo {volume} {127}},\ \bibinfo {pages} {241103} (\bibinfo {year} {2021})}\BibitemShut {NoStop}%
\bibitem [{\citenamefont {Gabbard}\ \emph {et~al.}(2021)\citenamefont {Gabbard}, \citenamefont {Messenger}, \citenamefont {Heng}, \citenamefont {Tonolini},\ and\ \citenamefont {Murray-Smith}}]{Gabbard2021}%
  \BibitemOpen
  \bibfield  {author} {\bibinfo {author} {\bibfnamefont {H.}~\bibnamefont {Gabbard}}, \bibinfo {author} {\bibfnamefont {C.}~\bibnamefont {Messenger}}, \bibinfo {author} {\bibfnamefont {I.~S.}\ \bibnamefont {Heng}}, \bibinfo {author} {\bibfnamefont {F.}~\bibnamefont {Tonolini}}, \ and\ \bibinfo {author} {\bibfnamefont {R.}~\bibnamefont {Murray-Smith}},\ }\href {\doibase 10.1038/s41567-021-01425-7} {\bibfield  {journal} {\bibinfo  {journal} {Nature Physics}\ }\textbf {\bibinfo {volume} {18}},\ \bibinfo {pages} {112–117} (\bibinfo {year} {2021})}\BibitemShut {NoStop}%
\bibitem [{\citenamefont {Islam}\ \emph {et~al.}(2022)\citenamefont {Islam}, \citenamefont {Roulet},\ and\ \citenamefont {Venumadhav}}]{Islam2022}%
  \BibitemOpen
  \bibfield  {author} {\bibinfo {author} {\bibfnamefont {T.}~\bibnamefont {Islam}}, \bibinfo {author} {\bibfnamefont {J.}~\bibnamefont {Roulet}}, \ and\ \bibinfo {author} {\bibfnamefont {T.}~\bibnamefont {Venumadhav}},\ }\href@noop {} {\enquote {\bibinfo {title} {Factorized parameter estimation for real-time gravitational wave inference},}\ } (\bibinfo {year} {2022}),\ \Eprint {http://arxiv.org/abs/2210.16278} {arXiv:2210.16278 [gr-qc]} \BibitemShut {NoStop}%
\bibitem [{\citenamefont {Rose}\ \emph {et~al.}(2022)\citenamefont {Rose}, \citenamefont {Valsan}, \citenamefont {Brady}, \citenamefont {Walsh},\ and\ \citenamefont {Pankow}}]{Rose2022}%
  \BibitemOpen
  \bibfield  {author} {\bibinfo {author} {\bibfnamefont {C.~A.}\ \bibnamefont {Rose}}, \bibinfo {author} {\bibfnamefont {V.}~\bibnamefont {Valsan}}, \bibinfo {author} {\bibfnamefont {P.~R.}\ \bibnamefont {Brady}}, \bibinfo {author} {\bibfnamefont {S.}~\bibnamefont {Walsh}}, \ and\ \bibinfo {author} {\bibfnamefont {C.}~\bibnamefont {Pankow}},\ }\href@noop {} {\enquote {\bibinfo {title} {Supplementing rapid {B}ayesian parameter estimation schemes with adaptive grids},}\ } (\bibinfo {year} {2022}),\ \Eprint {http://arxiv.org/abs/2201.05263} {arXiv:2201.05263 [gr-qc]} \BibitemShut {NoStop}%
\bibitem [{\citenamefont {Fairhurst}\ \emph {et~al.}(2023)\citenamefont {Fairhurst}, \citenamefont {Hoy}, \citenamefont {Green}, \citenamefont {Mills},\ and\ \citenamefont {Usman}}]{Fairhurst2023}%
  \BibitemOpen
  \bibfield  {author} {\bibinfo {author} {\bibfnamefont {S.}~\bibnamefont {Fairhurst}}, \bibinfo {author} {\bibfnamefont {C.}~\bibnamefont {Hoy}}, \bibinfo {author} {\bibfnamefont {R.}~\bibnamefont {Green}}, \bibinfo {author} {\bibfnamefont {C.}~\bibnamefont {Mills}}, \ and\ \bibinfo {author} {\bibfnamefont {S.~A.}\ \bibnamefont {Usman}},\ }\href {\doibase 10.1103/physrevd.108.082006} {\bibfield  {journal} {\bibinfo  {journal} {Physical Review D}\ }\textbf {\bibinfo {volume} {108}} (\bibinfo {year} {2023}),\ 10.1103/physrevd.108.082006}\BibitemShut {NoStop}%
\bibitem [{\citenamefont {Wong}\ \emph {et~al.}(2023)\citenamefont {Wong}, \citenamefont {Isi},\ and\ \citenamefont {Edwards}}]{Wong2023}%
  \BibitemOpen
  \bibfield  {author} {\bibinfo {author} {\bibfnamefont {K.~W.~K.}\ \bibnamefont {Wong}}, \bibinfo {author} {\bibfnamefont {M.}~\bibnamefont {Isi}}, \ and\ \bibinfo {author} {\bibfnamefont {T.~D.~P.}\ \bibnamefont {Edwards}},\ }\href {\doibase 10.3847/1538-4357/acf5cd} {\bibfield  {journal} {\bibinfo  {journal} {The Astrophysical Journal}\ }\textbf {\bibinfo {volume} {958}},\ \bibinfo {pages} {129} (\bibinfo {year} {2023})}\BibitemShut {NoStop}%
\bibitem [{\citenamefont {Roulet}\ \emph {et~al.}(2024)\citenamefont {Roulet}, \citenamefont {Mushkin}, \citenamefont {Wadekar}, \citenamefont {Venumadhav}, \citenamefont {Zackay},\ and\ \citenamefont {Zaldarriaga}}]{roulet2024extrinsic}%
  \BibitemOpen
  \bibfield  {author} {\bibinfo {author} {\bibfnamefont {J.}~\bibnamefont {Roulet}}, \bibinfo {author} {\bibfnamefont {J.}~\bibnamefont {Mushkin}}, \bibinfo {author} {\bibfnamefont {D.}~\bibnamefont {Wadekar}}, \bibinfo {author} {\bibfnamefont {T.}~\bibnamefont {Venumadhav}}, \bibinfo {author} {\bibfnamefont {B.}~\bibnamefont {Zackay}}, \ and\ \bibinfo {author} {\bibfnamefont {M.}~\bibnamefont {Zaldarriaga}},\ }\href {\doibase 10.1103/PhysRevD.110.044010} {\bibfield  {journal} {\bibinfo  {journal} {Phys. Rev. D}\ }\textbf {\bibinfo {volume} {110}},\ \bibinfo {pages} {044010} (\bibinfo {year} {2024})}\BibitemShut {NoStop}%
\bibitem [{\citenamefont {{Nitz}}(2024)}]{Nitz2024arXiv}%
  \BibitemOpen
  \bibfield  {author} {\bibinfo {author} {\bibfnamefont {A.~H.}\ \bibnamefont {{Nitz}}},\ }\href {\doibase 10.48550/arXiv.2410.05190} {\bibfield  {journal} {\bibinfo  {journal} {arXiv e-prints}\ ,\ \bibinfo {eid} {arXiv:2410.05190}} (\bibinfo {year} {2024})},\ \Eprint {http://arxiv.org/abs/2410.05190} {arXiv:2410.05190 [astro-ph.IM]} \BibitemShut {NoStop}%
\bibitem [{\citenamefont {Roulet}\ and\ \citenamefont {Venumadhav}(2024)}]{Roulet2024}%
  \BibitemOpen
  \bibfield  {author} {\bibinfo {author} {\bibfnamefont {J.}~\bibnamefont {Roulet}}\ and\ \bibinfo {author} {\bibfnamefont {T.}~\bibnamefont {Venumadhav}},\ }\href {\doibase 10.1146/annurev-nucl-121423-100725} {\bibfield  {journal} {\bibinfo  {journal} {Annual Review of Nuclear and Particle Science}\ }\textbf {\bibinfo {volume} {74}},\ \bibinfo {pages} {207–332} (\bibinfo {year} {2024})}\BibitemShut {NoStop}%
\bibitem [{\citenamefont {Allen}\ \emph {et~al.}(2012)\citenamefont {Allen}, \citenamefont {Anderson}, \citenamefont {Brady}, \citenamefont {Brown},\ and\ \citenamefont {Creighton}}]{Allen2012}%
  \BibitemOpen
  \bibfield  {author} {\bibinfo {author} {\bibfnamefont {B.}~\bibnamefont {Allen}}, \bibinfo {author} {\bibfnamefont {W.~G.}\ \bibnamefont {Anderson}}, \bibinfo {author} {\bibfnamefont {P.~R.}\ \bibnamefont {Brady}}, \bibinfo {author} {\bibfnamefont {D.~A.}\ \bibnamefont {Brown}}, \ and\ \bibinfo {author} {\bibfnamefont {J.~D.~E.}\ \bibnamefont {Creighton}},\ }\href {\doibase 10.1103/PhysRevD.85.122006} {\bibfield  {journal} {\bibinfo  {journal} {Phys. Rev. D}\ }\textbf {\bibinfo {volume} {85}},\ \bibinfo {pages} {122006} (\bibinfo {year} {2012})}\BibitemShut {NoStop}%
\bibitem [{\citenamefont {Usman}\ \emph {et~al.}(2016)\citenamefont {Usman}, \citenamefont {Nitz}, \citenamefont {Harry}, \citenamefont {Biwer}, \citenamefont {Brown}, \citenamefont {Cabero}, \citenamefont {Capano}, \citenamefont {Canton}, \citenamefont {Dent}, \citenamefont {Fairhurst}, \citenamefont {Kehl}, \citenamefont {Keppel}, \citenamefont {Krishnan}, \citenamefont {Lenon}, \citenamefont {Lundgren}, \citenamefont {Nielsen}, \citenamefont {Pekowsky}, \citenamefont {Pfeiffer}, \citenamefont {Saulson}, \citenamefont {West},\ and\ \citenamefont {Willis}}]{Usman2016}%
  \BibitemOpen
  \bibfield  {author} {\bibinfo {author} {\bibfnamefont {S.~A.}\ \bibnamefont {Usman}}, \bibinfo {author} {\bibfnamefont {A.~H.}\ \bibnamefont {Nitz}}, \bibinfo {author} {\bibfnamefont {I.~W.}\ \bibnamefont {Harry}}, \bibinfo {author} {\bibfnamefont {C.~M.}\ \bibnamefont {Biwer}},  \emph {et~al.},\ }\href {\doibase 10.1088/0264-9381/33/21/215004} {\bibfield  {journal} {\bibinfo  {journal} {Classical and Quantum Gravity}\ }\textbf {\bibinfo {volume} {33}},\ \bibinfo {pages} {215004} (\bibinfo {year} {2016})}\BibitemShut {NoStop}%
\bibitem [{\citenamefont {Messick}\ \emph {et~al.}(2017)\citenamefont {Messick}, \citenamefont {Blackburn}, \citenamefont {Brady}, \citenamefont {Brockill}, \citenamefont {Cannon}, \citenamefont {Cariou}, \citenamefont {Caudill}, \citenamefont {Chamberlin}, \citenamefont {Creighton}, \citenamefont {Everett}, \citenamefont {Hanna}, \citenamefont {Keppel}, \citenamefont {Lang}, \citenamefont {Li}, \citenamefont {Meacher}, \citenamefont {Nielsen}, \citenamefont {Pankow}, \citenamefont {Privitera}, \citenamefont {Qi}, \citenamefont {Sachdev}, \citenamefont {Sadeghian}, \citenamefont {Singer}, \citenamefont {Thomas}, \citenamefont {Wade}, \citenamefont {Wade}, \citenamefont {Weinstein},\ and\ \citenamefont {Wiesner}}]{Messick2017}%
  \BibitemOpen
  \bibfield  {author} {\bibinfo {author} {\bibfnamefont {C.}~\bibnamefont {Messick}}, \bibinfo {author} {\bibfnamefont {K.}~\bibnamefont {Blackburn}}, \bibinfo {author} {\bibfnamefont {P.}~\bibnamefont {Brady}}, \bibinfo {author} {\bibfnamefont {P.}~\bibnamefont {Brockill}},  \emph {et~al.},\ }\href {\doibase 10.1103/PhysRevD.95.042001} {\bibfield  {journal} {\bibinfo  {journal} {Phys. Rev. D}\ }\textbf {\bibinfo {volume} {95}},\ \bibinfo {pages} {042001} (\bibinfo {year} {2017})}\BibitemShut {NoStop}%
\bibitem [{\citenamefont {Venumadhav}\ \emph {et~al.}(2019)\citenamefont {Venumadhav}, \citenamefont {Zackay}, \citenamefont {Roulet}, \citenamefont {Dai},\ and\ \citenamefont {Zaldarriaga}}]{Venumadhav2019}%
  \BibitemOpen
  \bibfield  {author} {\bibinfo {author} {\bibfnamefont {T.}~\bibnamefont {Venumadhav}}, \bibinfo {author} {\bibfnamefont {B.}~\bibnamefont {Zackay}}, \bibinfo {author} {\bibfnamefont {J.}~\bibnamefont {Roulet}}, \bibinfo {author} {\bibfnamefont {L.}~\bibnamefont {Dai}}, \ and\ \bibinfo {author} {\bibfnamefont {M.}~\bibnamefont {Zaldarriaga}},\ }\href {\doibase 10.1103/PhysRevD.100.023011} {\bibfield  {journal} {\bibinfo  {journal} {Phys. Rev. D}\ }\textbf {\bibinfo {volume} {100}},\ \bibinfo {pages} {023011} (\bibinfo {year} {2019})}\BibitemShut {NoStop}%
\bibitem [{\citenamefont {{Neyman}}\ and\ \citenamefont {{Pearson}}(1933)}]{NP_lemma}%
  \BibitemOpen
  \bibfield  {author} {\bibinfo {author} {\bibfnamefont {J.}~\bibnamefont {{Neyman}}}\ and\ \bibinfo {author} {\bibfnamefont {E.~S.}\ \bibnamefont {{Pearson}}},\ }\href {\doibase 10.1098/rsta.1933.0009} {\bibfield  {journal} {\bibinfo  {journal} {Philosophical Transactions of the Royal Society of London Series A}\ }\textbf {\bibinfo {volume} {231}},\ \bibinfo {pages} {289} (\bibinfo {year} {1933})}\BibitemShut {NoStop}%
\bibitem [{\citenamefont {Allen}(2005)}]{Allen2005PhysRevD}%
  \BibitemOpen
  \bibfield  {author} {\bibinfo {author} {\bibfnamefont {B.}~\bibnamefont {Allen}},\ }\href {\doibase 10.1103/PhysRevD.71.062001} {\bibfield  {journal} {\bibinfo  {journal} {Phys. Rev. D}\ }\textbf {\bibinfo {volume} {71}},\ \bibinfo {pages} {062001} (\bibinfo {year} {2005})}\BibitemShut {NoStop}%
\bibitem [{\citenamefont {Owen}(2013)}]{Owen2013}%
  \BibitemOpen
  \bibfield  {author} {\bibinfo {author} {\bibfnamefont {A.~B.}\ \bibnamefont {Owen}},\ }\href@noop {} {\emph {\bibinfo {title} {Monte Carlo theory, methods and examples}}}\ (\bibinfo  {publisher} {\url{https://artowen.su.domains/mc/}},\ \bibinfo {year} {2013})\BibitemShut {NoStop}%
\bibitem [{\citenamefont {Morokoff}\ and\ \citenamefont {Caflisch}(1995)}]{Morokoff1995}%
  \BibitemOpen
  \bibfield  {author} {\bibinfo {author} {\bibfnamefont {W.~J.}\ \bibnamefont {Morokoff}}\ and\ \bibinfo {author} {\bibfnamefont {R.~E.}\ \bibnamefont {Caflisch}},\ }\href {\doibase https://doi.org/10.1006/jcph.1995.1209} {\bibfield  {journal} {\bibinfo  {journal} {Journal of Computational Physics}\ }\textbf {\bibinfo {volume} {122}},\ \bibinfo {pages} {218} (\bibinfo {year} {1995})}\BibitemShut {NoStop}%
\bibitem [{\citenamefont {Kong}(1992)}]{kong1992note}%
  \BibitemOpen
  \bibfield  {author} {\bibinfo {author} {\bibfnamefont {A.}~\bibnamefont {Kong}},\ }\href@noop {} {\bibfield  {journal} {\bibinfo  {journal} {University of Chicago, Dept. of Statistics, Tech. Rep}\ }\textbf {\bibinfo {volume} {348}},\ \bibinfo {pages} {14} (\bibinfo {year} {1992})}\BibitemShut {NoStop}%
\bibitem [{\citenamefont {Liu}\ and\ \citenamefont {Chen}(1995)}]{liu1995blind}%
  \BibitemOpen
  \bibfield  {author} {\bibinfo {author} {\bibfnamefont {J.~S.}\ \bibnamefont {Liu}}\ and\ \bibinfo {author} {\bibfnamefont {R.}~\bibnamefont {Chen}},\ }\href {https://doi.org/10.1080/01621459.1995.10476549} {\bibfield  {journal} {\bibinfo  {journal} {Journal of the american statistical association}\ }\textbf {\bibinfo {volume} {90}},\ \bibinfo {pages} {567} (\bibinfo {year} {1995})}\BibitemShut {NoStop}%
\bibitem [{\citenamefont {{Metropolis}}\ \emph {et~al.}(1953)\citenamefont {{Metropolis}}, \citenamefont {{Rosenbluth}}, \citenamefont {{Rosenbluth}}, \citenamefont {{Teller}},\ and\ \citenamefont {{Teller}}}]{1953JChPh..21.1087M}%
  \BibitemOpen
  \bibfield  {author} {\bibinfo {author} {\bibfnamefont {N.}~\bibnamefont {{Metropolis}}}, \bibinfo {author} {\bibfnamefont {A.~W.}\ \bibnamefont {{Rosenbluth}}}, \bibinfo {author} {\bibfnamefont {M.~N.}\ \bibnamefont {{Rosenbluth}}}, \bibinfo {author} {\bibfnamefont {A.~H.}\ \bibnamefont {{Teller}}}, \ and\ \bibinfo {author} {\bibfnamefont {E.}~\bibnamefont {{Teller}}},\ }\href {\doibase 10.1063/1.1699114} {\bibfield  {journal} {\bibinfo  {journal} {\jcp}\ }\textbf {\bibinfo {volume} {21}},\ \bibinfo {pages} {1087} (\bibinfo {year} {1953})}\BibitemShut {NoStop}%
\bibitem [{\citenamefont {{Hastings}}(1970)}]{Hastings1970}%
  \BibitemOpen
  \bibfield  {author} {\bibinfo {author} {\bibfnamefont {W.~K.}\ \bibnamefont {{Hastings}}},\ }\href {\doibase 10.1093/biomet/57.1.97} {\bibfield  {journal} {\bibinfo  {journal} {Biometrika}\ }\textbf {\bibinfo {volume} {57}},\ \bibinfo {pages} {97} (\bibinfo {year} {1970})}\BibitemShut {NoStop}%
\bibitem [{\citenamefont {Skilling}(2006)}]{skilling2006nested}%
  \BibitemOpen
  \bibfield  {author} {\bibinfo {author} {\bibfnamefont {J.}~\bibnamefont {Skilling}},\ }\href {\doibase 10.1214/06-ba127} {\bibfield  {journal} {\bibinfo  {journal} {Bayesian Analysis}\ }\textbf {\bibinfo {volume} {1}} (\bibinfo {year} {2006}),\ 10.1214/06-ba127}\BibitemShut {NoStop}%
\bibitem [{\citenamefont {{Buchner}}(2023)}]{Buchner2023Nested}%
  \BibitemOpen
  \bibfield  {author} {\bibinfo {author} {\bibfnamefont {J.}~\bibnamefont {{Buchner}}},\ }\href {\doibase 10.1214/23-SS144} {\bibfield  {journal} {\bibinfo  {journal} {Statistics Surveys}\ }\textbf {\bibinfo {volume} {17}},\ \bibinfo {pages} {169} (\bibinfo {year} {2023})},\ \Eprint {http://arxiv.org/abs/2101.09675} {arXiv:2101.09675 [stat.CO]} \BibitemShut {NoStop}%
\bibitem [{\citenamefont {Gelman}\ and\ \citenamefont {Meng}(1998)}]{Gelman1998}%
  \BibitemOpen
  \bibfield  {author} {\bibinfo {author} {\bibfnamefont {A.}~\bibnamefont {Gelman}}\ and\ \bibinfo {author} {\bibfnamefont {X.-L.}\ \bibnamefont {Meng}},\ }\href {http://www.jstor.org/stable/2676756} {\bibfield  {journal} {\bibinfo  {journal} {Statistical Science}\ }\textbf {\bibinfo {volume} {13}},\ \bibinfo {pages} {163} (\bibinfo {year} {1998})}\BibitemShut {NoStop}%
\bibitem [{\citenamefont {{Speagle}}(2020)}]{Speagle2020}%
  \BibitemOpen
  \bibfield  {author} {\bibinfo {author} {\bibfnamefont {J.~S.}\ \bibnamefont {{Speagle}}},\ }\href {\doibase 10.1093/mnras/staa278} {\bibfield  {journal} {\bibinfo  {journal} {Monthly Notices of the Royal Astronomical Society}\ }\textbf {\bibinfo {volume} {493}},\ \bibinfo {pages} {3132} (\bibinfo {year} {2020})},\ \Eprint {http://arxiv.org/abs/1904.02180} {arXiv:1904.02180 [astro-ph.IM]} \BibitemShut {NoStop}%
\bibitem [{\citenamefont {{Ashton}}\ \emph {et~al.}(2022)\citenamefont {{Ashton}}, \citenamefont {{Bernstein}}, \citenamefont {{Buchner}}, \citenamefont {{Chen}}, \citenamefont {{Cs{\'a}nyi}}, \citenamefont {{Fowlie}}, \citenamefont {{Feroz}}, \citenamefont {{Griffiths}}, \citenamefont {{Handley}}, \citenamefont {{Habeck}}, \citenamefont {{Higson}}, \citenamefont {{Hobson}}, \citenamefont {{Lasenby}}, \citenamefont {{Parkinson}}, \citenamefont {{P{\'a}rtay}}, \citenamefont {{Pitkin}}, \citenamefont {{Schneider}}, \citenamefont {{Speagle}}, \citenamefont {{South}}, \citenamefont {{Veitch}}, \citenamefont {{Wacker}}, \citenamefont {{Wales}},\ and\ \citenamefont {{Yallup}}}]{Ashton2022}%
  \BibitemOpen
  \bibfield  {author} {\bibinfo {author} {\bibfnamefont {G.}~\bibnamefont {{Ashton}}}, \bibinfo {author} {\bibfnamefont {N.}~\bibnamefont {{Bernstein}}}, \bibinfo {author} {\bibfnamefont {J.}~\bibnamefont {{Buchner}}}, \bibinfo {author} {\bibfnamefont {X.}~\bibnamefont {{Chen}}},  \emph {et~al.},\ }\href {\doibase 10.1038/s43586-022-00121-x} {\bibfield  {journal} {\bibinfo  {journal} {Nature Reviews Methods Primers}\ }\textbf {\bibinfo {volume} {2}},\ \bibinfo {eid} {39} (\bibinfo {year} {2022})},\ \Eprint {http://arxiv.org/abs/2205.15570} {arXiv:2205.15570 [stat.CO]} \BibitemShut {NoStop}%
\bibitem [{\citenamefont {{Apostolatos}}\ \emph {et~al.}(1994)\citenamefont {{Apostolatos}}, \citenamefont {{Cutler}}, \citenamefont {{Sussman}},\ and\ \citenamefont {{Thorne}}}]{Apostolatos1994}%
  \BibitemOpen
  \bibfield  {author} {\bibinfo {author} {\bibfnamefont {T.~A.}\ \bibnamefont {{Apostolatos}}}, \bibinfo {author} {\bibfnamefont {C.}~\bibnamefont {{Cutler}}}, \bibinfo {author} {\bibfnamefont {G.~J.}\ \bibnamefont {{Sussman}}}, \ and\ \bibinfo {author} {\bibfnamefont {K.~S.}\ \bibnamefont {{Thorne}}},\ }\href {\doibase 10.1103/PhysRevD.49.6274} {\bibfield  {journal} {\bibinfo  {journal} {\prd}\ }\textbf {\bibinfo {volume} {49}},\ \bibinfo {pages} {6274} (\bibinfo {year} {1994})}\BibitemShut {NoStop}%
\bibitem [{\citenamefont {Schmidt}\ \emph {et~al.}(2012)\citenamefont {Schmidt}, \citenamefont {Hannam},\ and\ \citenamefont {Husa}}]{Schmidt2012}%
  \BibitemOpen
  \bibfield  {author} {\bibinfo {author} {\bibfnamefont {P.}~\bibnamefont {Schmidt}}, \bibinfo {author} {\bibfnamefont {M.}~\bibnamefont {Hannam}}, \ and\ \bibinfo {author} {\bibfnamefont {S.}~\bibnamefont {Husa}},\ }\href {\doibase 10.1103/PhysRevD.86.104063} {\bibfield  {journal} {\bibinfo  {journal} {Phys. Rev. D}\ }\textbf {\bibinfo {volume} {86}},\ \bibinfo {pages} {104063} (\bibinfo {year} {2012})}\BibitemShut {NoStop}%
\bibitem [{\citenamefont {Pratten}\ \emph {et~al.}(2021)\citenamefont {Pratten}, \citenamefont {Garc\'{\i}a-Quir\'os}, \citenamefont {Colleoni}, \citenamefont {Ramos-Buades}, \citenamefont {Estell\'es}, \citenamefont {Mateu-Lucena}, \citenamefont {Jaume}, \citenamefont {Haney}, \citenamefont {Keitel}, \citenamefont {Thompson},\ and\ \citenamefont {Husa}}]{Pratten2021}%
  \BibitemOpen
  \bibfield  {author} {\bibinfo {author} {\bibfnamefont {G.}~\bibnamefont {Pratten}}, \bibinfo {author} {\bibfnamefont {C.}~\bibnamefont {Garc\'{\i}a-Quir\'os}}, \bibinfo {author} {\bibfnamefont {M.}~\bibnamefont {Colleoni}}, \bibinfo {author} {\bibfnamefont {A.}~\bibnamefont {Ramos-Buades}},  \emph {et~al.},\ }\href {\doibase 10.1103/PhysRevD.103.104056} {\bibfield  {journal} {\bibinfo  {journal} {Phys. Rev. D}\ }\textbf {\bibinfo {volume} {103}},\ \bibinfo {pages} {104056} (\bibinfo {year} {2021})}\BibitemShut {NoStop}%
\bibitem [{\citenamefont {Yu}\ \emph {et~al.}(2023)\citenamefont {Yu}, \citenamefont {Roulet}, \citenamefont {Venumadhav}, \citenamefont {Zackay},\ and\ \citenamefont {Zaldarriaga}}]{Yu2023}%
  \BibitemOpen
  \bibfield  {author} {\bibinfo {author} {\bibfnamefont {H.}~\bibnamefont {Yu}}, \bibinfo {author} {\bibfnamefont {J.}~\bibnamefont {Roulet}}, \bibinfo {author} {\bibfnamefont {T.}~\bibnamefont {Venumadhav}}, \bibinfo {author} {\bibfnamefont {B.}~\bibnamefont {Zackay}}, \ and\ \bibinfo {author} {\bibfnamefont {M.}~\bibnamefont {Zaldarriaga}},\ }\href {\doibase 10.1103/PhysRevD.108.064059} {\bibfield  {journal} {\bibinfo  {journal} {Phys. Rev. D}\ }\textbf {\bibinfo {volume} {108}},\ \bibinfo {pages} {064059} (\bibinfo {year} {2023})}\BibitemShut {NoStop}%
\bibitem [{\citenamefont {Ossokine}\ \emph {et~al.}(2020)\citenamefont {Ossokine}, \citenamefont {Buonanno}, \citenamefont {Marsat}, \citenamefont {Cotesta}, \citenamefont {Babak}, \citenamefont {Dietrich}, \citenamefont {Haas}, \citenamefont {Hinder}, \citenamefont {Pfeiffer}, \citenamefont {P\"urrer}, \citenamefont {Woodford}, \citenamefont {Boyle}, \citenamefont {Kidder}, \citenamefont {Scheel},\ and\ \citenamefont {Szil\'agyi}}]{Ossokine2020PhysRevD}%
  \BibitemOpen
  \bibfield  {author} {\bibinfo {author} {\bibfnamefont {S.}~\bibnamefont {Ossokine}}, \bibinfo {author} {\bibfnamefont {A.}~\bibnamefont {Buonanno}}, \bibinfo {author} {\bibfnamefont {S.}~\bibnamefont {Marsat}}, \bibinfo {author} {\bibfnamefont {R.}~\bibnamefont {Cotesta}},  \emph {et~al.},\ }\href {\doibase 10.1103/PhysRevD.102.044055} {\bibfield  {journal} {\bibinfo  {journal} {Phys. Rev. D}\ }\textbf {\bibinfo {volume} {102}},\ \bibinfo {pages} {044055} (\bibinfo {year} {2020})}\BibitemShut {NoStop}%
\bibitem [{\citenamefont {Whelan}(2013)}]{Whelan2013}%
  \BibitemOpen
  \bibfield  {author} {\bibinfo {author} {\bibfnamefont {J.~T.}\ \bibnamefont {Whelan}},\ }\href@noop {} {\enquote {\bibinfo {title} {The geometry of gravitational wave detection},}\ }\bibinfo {howpublished} {\url{https://dcc.ligo.org/public/0106/T1300666/003/Whelan_geometry.pdf}} (\bibinfo {year} {2013})\BibitemShut {NoStop}%
\bibitem [{\citenamefont {Singer}\ and\ \citenamefont {Price}(2016)}]{Singer2016}%
  \BibitemOpen
  \bibfield  {author} {\bibinfo {author} {\bibfnamefont {L.~P.}\ \bibnamefont {Singer}}\ and\ \bibinfo {author} {\bibfnamefont {L.~R.}\ \bibnamefont {Price}},\ }\href {\doibase 10.1103/PhysRevD.93.024013} {\bibfield  {journal} {\bibinfo  {journal} {Phys. Rev. D}\ }\textbf {\bibinfo {volume} {93}},\ \bibinfo {pages} {024013} (\bibinfo {year} {2016})}\BibitemShut {NoStop}%
\bibitem [{\citenamefont {Tiwari}\ \emph {et~al.}(2023)\citenamefont {Tiwari}, \citenamefont {Hoy}, \citenamefont {Fairhurst},\ and\ \citenamefont {MacLeod}}]{Tiwari2023PhysRevD}%
  \BibitemOpen
  \bibfield  {author} {\bibinfo {author} {\bibfnamefont {V.}~\bibnamefont {Tiwari}}, \bibinfo {author} {\bibfnamefont {C.}~\bibnamefont {Hoy}}, \bibinfo {author} {\bibfnamefont {S.}~\bibnamefont {Fairhurst}}, \ and\ \bibinfo {author} {\bibfnamefont {D.}~\bibnamefont {MacLeod}},\ }\href {\doibase 10.1103/PhysRevD.108.023001} {\bibfield  {journal} {\bibinfo  {journal} {Phys. Rev. D}\ }\textbf {\bibinfo {volume} {108}},\ \bibinfo {pages} {023001} (\bibinfo {year} {2023})}\BibitemShut {NoStop}%
\bibitem [{\citenamefont {{Chan}}\ \emph {et~al.}(2018)\citenamefont {{Chan}}, \citenamefont {{Messenger}}, \citenamefont {{Heng}},\ and\ \citenamefont {{Hendry}}}]{Chan2018PhRvD}%
  \BibitemOpen
  \bibfield  {author} {\bibinfo {author} {\bibfnamefont {M.~L.}\ \bibnamefont {{Chan}}}, \bibinfo {author} {\bibfnamefont {C.}~\bibnamefont {{Messenger}}}, \bibinfo {author} {\bibfnamefont {I.~S.}\ \bibnamefont {{Heng}}}, \ and\ \bibinfo {author} {\bibfnamefont {M.}~\bibnamefont {{Hendry}}},\ }\href {\doibase 10.1103/PhysRevD.97.123014} {\bibfield  {journal} {\bibinfo  {journal} {\prd}\ }\textbf {\bibinfo {volume} {97}},\ \bibinfo {eid} {123014} (\bibinfo {year} {2018})},\ \Eprint {http://arxiv.org/abs/1803.09680} {arXiv:1803.09680 [astro-ph.HE]} \BibitemShut {NoStop}%
\bibitem [{\citenamefont {Strassen}(1969)}]{strassen1969gaussian}%
  \BibitemOpen
  \bibfield  {author} {\bibinfo {author} {\bibfnamefont {V.}~\bibnamefont {Strassen}},\ }\href@noop {} {\bibfield  {journal} {\bibinfo  {journal} {Numerische mathematik}\ }\textbf {\bibinfo {volume} {13}},\ \bibinfo {pages} {354} (\bibinfo {year} {1969})}\BibitemShut {NoStop}%
\bibitem [{\citenamefont {Bj{\"o}rck}\ \emph {et~al.}(2015)\citenamefont {Bj{\"o}rck} \emph {et~al.}}]{bjorck2015numerical}%
  \BibitemOpen
  \bibfield  {author} {\bibinfo {author} {\bibfnamefont {{\AA}.}~\bibnamefont {Bj{\"o}rck}} \emph {et~al.},\ }\href@noop {} {\emph {\bibinfo {title} {Numerical methods in matrix computations}}},\ Vol.~\bibinfo {volume} {59}\ (\bibinfo  {publisher} {Springer},\ \bibinfo {year} {2015})\BibitemShut {NoStop}%
\bibitem [{\citenamefont {Vallisneri}(2008)}]{Vallisneri2008PhysRevD}%
  \BibitemOpen
  \bibfield  {author} {\bibinfo {author} {\bibfnamefont {M.}~\bibnamefont {Vallisneri}},\ }\href {\doibase 10.1103/PhysRevD.77.042001} {\bibfield  {journal} {\bibinfo  {journal} {Phys. Rev. D}\ }\textbf {\bibinfo {volume} {77}},\ \bibinfo {pages} {042001} (\bibinfo {year} {2008})}\BibitemShut {NoStop}%
\bibitem [{\citenamefont {{Pratten}}\ \emph {et~al.}(2020)\citenamefont {{Pratten}}, \citenamefont {{Husa}}, \citenamefont {{Garc{\'\i}a-Quir{\'o}s}}, \citenamefont {{Colleoni}}, \citenamefont {{Ramos-Buades}}, \citenamefont {{Estell{\'e}s}},\ and\ \citenamefont {{Jaume}}}]{Pratten2020PhRvD}%
  \BibitemOpen
  \bibfield  {author} {\bibinfo {author} {\bibfnamefont {G.}~\bibnamefont {{Pratten}}}, \bibinfo {author} {\bibfnamefont {S.}~\bibnamefont {{Husa}}}, \bibinfo {author} {\bibfnamefont {C.}~\bibnamefont {{Garc{\'\i}a-Quir{\'o}s}}}, \bibinfo {author} {\bibfnamefont {M.}~\bibnamefont {{Colleoni}}}, \bibinfo {author} {\bibfnamefont {A.}~\bibnamefont {{Ramos-Buades}}}, \bibinfo {author} {\bibfnamefont {H.}~\bibnamefont {{Estell{\'e}s}}}, \ and\ \bibinfo {author} {\bibfnamefont {R.}~\bibnamefont {{Jaume}}},\ }\href {\doibase 10.1103/PhysRevD.102.064001} {\bibfield  {journal} {\bibinfo  {journal} {\prd}\ }\textbf {\bibinfo {volume} {102}},\ \bibinfo {eid} {064001} (\bibinfo {year} {2020})},\ \Eprint {http://arxiv.org/abs/2001.11412} {arXiv:2001.11412 [gr-qc]} \BibitemShut {NoStop}%
\bibitem [{\citenamefont {{Cutler}}\ and\ \citenamefont {{Flanagan}}(1994)}]{Cutler1994PhRvD}%
  \BibitemOpen
  \bibfield  {author} {\bibinfo {author} {\bibfnamefont {C.}~\bibnamefont {{Cutler}}}\ and\ \bibinfo {author} {\bibfnamefont {{\'E}.~E.}\ \bibnamefont {{Flanagan}}},\ }\href {\doibase 10.1103/PhysRevD.49.2658} {\bibfield  {journal} {\bibinfo  {journal} {\prd}\ }\textbf {\bibinfo {volume} {49}},\ \bibinfo {pages} {2658} (\bibinfo {year} {1994})},\ \Eprint {http://arxiv.org/abs/gr-qc/9402014} {arXiv:gr-qc/9402014 [gr-qc]} \BibitemShut {NoStop}%
\bibitem [{\citenamefont {Roulet}\ \emph {et~al.}(2022)\citenamefont {Roulet}, \citenamefont {Olsen}, \citenamefont {Mushkin}, \citenamefont {Islam}, \citenamefont {Venumadhav}, \citenamefont {Zackay},\ and\ \citenamefont {Zaldarriaga}}]{Roulet2022}%
  \BibitemOpen
  \bibfield  {author} {\bibinfo {author} {\bibfnamefont {J.}~\bibnamefont {Roulet}}, \bibinfo {author} {\bibfnamefont {S.}~\bibnamefont {Olsen}}, \bibinfo {author} {\bibfnamefont {J.}~\bibnamefont {Mushkin}}, \bibinfo {author} {\bibfnamefont {T.}~\bibnamefont {Islam}}, \bibinfo {author} {\bibfnamefont {T.}~\bibnamefont {Venumadhav}}, \bibinfo {author} {\bibfnamefont {B.}~\bibnamefont {Zackay}}, \ and\ \bibinfo {author} {\bibfnamefont {M.}~\bibnamefont {Zaldarriaga}},\ }\href {\doibase 10.1103/PhysRevD.106.123015} {\bibfield  {journal} {\bibinfo  {journal} {Phys. Rev. D}\ }\textbf {\bibinfo {volume} {106}},\ \bibinfo {pages} {123015} (\bibinfo {year} {2022})}\BibitemShut {NoStop}%
\bibitem [{\citenamefont {Singer}\ \emph {et~al.}(2014)\citenamefont {Singer}, \citenamefont {Price}, \citenamefont {Farr}, \citenamefont {Urban}, \citenamefont {Pankow}, \citenamefont {Vitale}, \citenamefont {Veitch}, \citenamefont {Farr}, \citenamefont {Hanna}, \citenamefont {Cannon}, \citenamefont {Downes}, \citenamefont {Graff}, \citenamefont {Haster}, \citenamefont {Mandel}, \citenamefont {Sidery},\ and\ \citenamefont {Vecchio}}]{Singer2014}%
  \BibitemOpen
  \bibfield  {author} {\bibinfo {author} {\bibfnamefont {L.~P.}\ \bibnamefont {Singer}}, \bibinfo {author} {\bibfnamefont {L.~R.}\ \bibnamefont {Price}}, \bibinfo {author} {\bibfnamefont {B.}~\bibnamefont {Farr}}, \bibinfo {author} {\bibfnamefont {A.~L.}\ \bibnamefont {Urban}},  \emph {et~al.},\ }\href {\doibase 10.1088/0004-637x/795/2/105} {\bibfield  {journal} {\bibinfo  {journal} {The Astrophysical Journal}\ }\textbf {\bibinfo {volume} {795}},\ \bibinfo {pages} {105} (\bibinfo {year} {2014})}\BibitemShut {NoStop}%
\bibitem [{\citenamefont {{Lange}}(2023)}]{Lange2023}%
  \BibitemOpen
  \bibfield  {author} {\bibinfo {author} {\bibfnamefont {J.~U.}\ \bibnamefont {{Lange}}},\ }\href {\doibase 10.1093/mnras/stad2441} {\bibfield  {journal} {\bibinfo  {journal} {Monthly Notices of the Royal Astronomical Society}\ }\textbf {\bibinfo {volume} {525}},\ \bibinfo {pages} {3181} (\bibinfo {year} {2023})},\ \Eprint {http://arxiv.org/abs/2306.16923} {arXiv:2306.16923 [astro-ph.IM]} \BibitemShut {NoStop}%
\bibitem [{\citenamefont {Cornish}(2013)}]{Cornish2013}%
  \BibitemOpen
  \bibfield  {author} {\bibinfo {author} {\bibfnamefont {N.~J.}\ \bibnamefont {Cornish}},\ }\href@noop {} {\enquote {\bibinfo {title} {Fast {F}isher matrices and lazy likelihoods},}\ } (\bibinfo {year} {2013}),\ \Eprint {http://arxiv.org/abs/1007.4820} {arXiv:1007.4820 [gr-qc]} \BibitemShut {NoStop}%
\bibitem [{\citenamefont {Leslie}\ \emph {et~al.}(2021)\citenamefont {Leslie}, \citenamefont {Dai},\ and\ \citenamefont {Pratten}}]{Leslie2021}%
  \BibitemOpen
  \bibfield  {author} {\bibinfo {author} {\bibfnamefont {N.}~\bibnamefont {Leslie}}, \bibinfo {author} {\bibfnamefont {L.}~\bibnamefont {Dai}}, \ and\ \bibinfo {author} {\bibfnamefont {G.}~\bibnamefont {Pratten}},\ }\href {\doibase 10.1103/PhysRevD.104.123030} {\bibfield  {journal} {\bibinfo  {journal} {Phys. Rev. D}\ }\textbf {\bibinfo {volume} {104}},\ \bibinfo {pages} {123030} (\bibinfo {year} {2021})}\BibitemShut {NoStop}%
\bibitem [{\citenamefont {{McIsaac}}\ \emph {et~al.}(2023)\citenamefont {{McIsaac}}, \citenamefont {{Hoy}},\ and\ \citenamefont {{Harry}}}]{2023PhRvD.108l3016M}%
  \BibitemOpen
  \bibfield  {author} {\bibinfo {author} {\bibfnamefont {C.}~\bibnamefont {{McIsaac}}}, \bibinfo {author} {\bibfnamefont {C.}~\bibnamefont {{Hoy}}}, \ and\ \bibinfo {author} {\bibfnamefont {I.}~\bibnamefont {{Harry}}},\ }\href {\doibase 10.1103/PhysRevD.108.123016} {\bibfield  {journal} {\bibinfo  {journal} {\prd}\ }\textbf {\bibinfo {volume} {108}},\ \bibinfo {eid} {123016} (\bibinfo {year} {2023})},\ \Eprint {http://arxiv.org/abs/2303.17364} {arXiv:2303.17364 [gr-qc]} \BibitemShut {NoStop}%
\bibitem [{\citenamefont {{Pan}}\ \emph {et~al.}(2004)\citenamefont {{Pan}}, \citenamefont {{Buonanno}}, \citenamefont {{Chen}},\ and\ \citenamefont {{Vallisneri}}}]{Pan2015PhysRvD}%
  \BibitemOpen
  \bibfield  {author} {\bibinfo {author} {\bibfnamefont {Y.}~\bibnamefont {{Pan}}}, \bibinfo {author} {\bibfnamefont {A.}~\bibnamefont {{Buonanno}}}, \bibinfo {author} {\bibfnamefont {Y.}~\bibnamefont {{Chen}}}, \ and\ \bibinfo {author} {\bibfnamefont {M.}~\bibnamefont {{Vallisneri}}},\ }\href {\doibase 10.1103/PhysRevD.69.104017} {\bibfield  {journal} {\bibinfo  {journal} {\prd}\ }\textbf {\bibinfo {volume} {69}},\ \bibinfo {eid} {104017} (\bibinfo {year} {2004})},\ \Eprint {http://arxiv.org/abs/gr-qc/0310034} {arXiv:gr-qc/0310034 [gr-qc]} \BibitemShut {NoStop}%
\bibitem [{\citenamefont {{Harry}}\ \emph {et~al.}(2016)\citenamefont {{Harry}}, \citenamefont {{Privitera}}, \citenamefont {{Boh{\'e}}},\ and\ \citenamefont {{Buonanno}}}]{2016PhRvD..94b4012H}%
  \BibitemOpen
  \bibfield  {author} {\bibinfo {author} {\bibfnamefont {I.}~\bibnamefont {{Harry}}}, \bibinfo {author} {\bibfnamefont {S.}~\bibnamefont {{Privitera}}}, \bibinfo {author} {\bibfnamefont {A.}~\bibnamefont {{Boh{\'e}}}}, \ and\ \bibinfo {author} {\bibfnamefont {A.}~\bibnamefont {{Buonanno}}},\ }\href {\doibase 10.1103/PhysRevD.94.024012} {\bibfield  {journal} {\bibinfo  {journal} {\prd}\ }\textbf {\bibinfo {volume} {94}},\ \bibinfo {eid} {024012} (\bibinfo {year} {2016})},\ \Eprint {http://arxiv.org/abs/1603.02444} {arXiv:1603.02444 [gr-qc]} \BibitemShut {NoStop}%
\end{thebibliography}%
\end{document}